\documentclass[10pt]{iopart}

\begin{document}

\title[Optimization of Synchronization in Noisy Weighted Complex Networks]
{Synchronization in Weighted Uncorrelated Complex Networks in a Noisy Environment: 
Optimization and Connections with Transport Efficiency}
\author{G. Korniss}

\address{Department of Physics, Applied Physics, and Astronomy, 
Rensselaer Polytechnic Institute, 110 8$^{th}$ Street, Troy, NY 12180--3590, USA}
\ead{korniss@rpi.edu}
\begin{abstract}
Motivated by synchronization problems in noisy environments, we
study the Edwards-Wilkinson process on weighted uncorrelated scale-free
networks. We consider a specific form of the weights, where the
strength (and the associated cost) of a link is proportional to
$(k_{i}k_{j})^{\beta}$ with $k_{i}$ and $k_{j}$ being the degrees
of the nodes connected by the link. Subject to the constraint that
the total network cost is fixed, we find that in the mean-field
approximation on uncorrelated scale-free graphs, synchronization
is optimal at $\beta^{*}$$=$$-1$. Numerical results, based on
exact numerical diagonalization of the corresponding network
Laplacian, confirm the mean-field results, with small corrections
to the optimal value of $\beta^{*}$. Employing our recent
connections between the Edwards-Wilkinson process and resistor
networks, and some well-known connections between random walks and
resistor networks, we also pursue a naturally related problem of
optimizing performance in queue-limited communication networks
utilizing local weighted routing schemes.
\end{abstract}
\vspace{2pc}
\maketitle

\section{Introduction}

Synchronization in natural and artificial complex interconnected
systems \cite{Strogatz_review,Strogatz_book,Boccaletti_review,
Acebron_review} has been the focus of interdisciplinary research with
applications ranging from neurobiology \cite{Netoff04,GRIN05},
ecology and population dynamics \cite{Winfree67,Lusseau06} to
scalable computer networks
\cite{Korniss_PRL2000,KORNISS03,Sloot01,KIRK2004,Alice_review}. 
In the recent flood of research on complex networks,
\cite{WATTS98,Barab_sci,BarabREV,MendesREV,NEWMAN_SIAM,Willinger_review},
the focus has shifted
from structure to various dynamical and stochastic processes on
networks, synchronization is being one of them. The common question
addressed by most studies within their specific context is how the
collective response of locally-coupled entities is influenced by the
underlying network topology.

A large number of studies investigated the Kuramoto model of coupled
oscillators \cite{Acebron_review,Kuramoto}, naturally
generalized to complex networks
\cite{Hong2002,Ichinomiya2004,Lee2005}.
The common feature of the findings is the spontaneous emergence of
order (synchronous phase) on complex networks, qualitatively similar
to that observed on fully connected networks, in contrast to regular networks in low dimensions.
Another large group of studies addressed synchronization in coupled
nonlinear dynamical systems (e.g., chaotic oscillators)
\cite{Boccaletti_review} on small-world (SW) \cite{BARAHONA02}  and
scale-free (SF) \cite{LAI03,MOTTER05a,MOTTER05b,ZHOU06a,ZHOU06b}
networks. The analysis of synchronization in the latter models can be
carried out by linearization about the synchronous state and using the framework of
the master stability function \cite{Pecora98}. In turn, the technical
challenge of the problem is reduced to the diagonalization of the
Laplacian on the respective network, and calculating or estimating the
eigenratio \cite{BARAHONA02} (the ratio of the largest and the smallest non-zero
eigenvalue of the network Laplacian), a characteristic measure of
synchronizability (smaller eigenratios imply better synchronizability).
Along these lines, most recent studies \cite{MOTTER05a,MOTTER05b,ZHOU06a,Donetti_JSM2006} 
considered not only complex, possibly heterogeneous, interaction topologies between the
nodes, but also heterogeneities in the strength of the couplings (also
referred to as weighted complex networks).

In a more general setting of synchronization problems, the collective
behavior/response of the system is obviously strongly influenced by the nonlinearities, the
coupling/interaction topology, the weights/strength of the (possibly directed) links,
and the presence and the type of noise
\cite{Boccaletti_review,ZHOU06b}. Here, we study synchronization in
weighted complex networks with linear coupling in the presence of
delta-correlated white noise. Although it may appear somewhat
simplistic (and, indeed prototypical), such problems are motivated by the
dynamics and fluctuations in task
completion landscapes in causally-constrained  queuing
networks \cite{TORO_VIRTUAL}, with applications in manufacturing
supply chains, e-commerce-based services facilitated by interconnected
servers \cite{Dong2005}, and certain distributed-computing schemes on
computer networks \cite{Korniss_PRL2000,KORNISS03,Sloot01,KIRK2004,Alice_review}. This simplified problem
is the Edwards-Wilkinson (EW) process \cite{EW} on the respective network
\cite{Korniss_PLA,KHK04,KHK05,KHK_SPIE_2005,ee}.

Here, we consider a specific and symmetric form of the weights on
uncorrelated SF networks, being proportional to
$(k_{i}k_{j})^{\beta}$ where $k_{i}$ and $k_{j}$ are the
degrees of the nodes connected by the link \cite{ZHOU06a}. The above
general form has been suggested by empirical studies of metabolic \cite{Macdonald} and
airline transportation networks \cite{barrat}. Here, we study the
effect of such weighting scheme in our synchronization problem.
Associating the weight/strength of each link with its cost, we ask
what is the optimal allocation of the weights (in terms of $\beta$) in
strongly heterogeneous networks, with a fixed total cost, in order to maximize
synchronization in a noisy environment. For the EW process on any
network, the natural observable is the width or spread of the
synchronization landscape
\cite{Korniss_PLA,KHK04,KHK05,KHK_SPIE_2005,ee}. Then the task
becomes minimizing the width as a function of $\beta$ subject to a (cost) constraint.

Despite differences in the assumptions concerning noise and constrained cost, 
our study's results are very
similar to findings by Zhou et al. \cite{ZHOU06a}, who investigated
synchronization of coupled nonlinear oscillators on the same type of
network. The optimal value of $\beta$ is
close to $-1$ (and is exactly  $-1$ in the mean-field approximation on
uncorrelated random SF networks.) The two problems are intimately
connected through the eigenvalue spectrum of the same network Laplacian.

Transport and flow on complex networks have also become the subject of
intensive research with applications to biological, transportation, communication, and infrastructure
networks
\cite{Donetti_JSM2006,barrat,Goh_PRL2001,Goh_PRE2005,Noh_PRL2004,BARA03,
Almaas,TORO04,Amaral_PNAS2005,LAI05,Ashton05,Rieger05,Brockmann_2006,
Vespignani_PNAS2006,Krause_PA04,Guimera_PRL04,Tadic_PRE2005,Zhao_05,Toro_PRL06,Danila_PRE06,
Danila_PRL06,Andrade_PRL2005,Lopez2005, Wu_PRL2005}.
While our main motivation is the above described
synchronization phenomena in noisy environments, we also explore some
natural connections
with idealized transport and flow problems on complex
networks, in particular, connections with local routing schemes
\cite{Guimera_PRL04,Tadic_PRE2005,Danila_PRE06}.
Connections between random walks and resistor networks have
been discussed in detail in several works \cite{Doyle,Lovasz,Redner}.
Further, we have recently pointed out \cite{Korniss_PLA} some useful connections between
the EW process and resistor networks (both systems' behavior is
governed by the same network Laplacian). Thus, our results for the
synchronization problem have some straightforward implications on the related
resistor network and random walk problems, pursued in the second part
of this work.

The remainder of the paper is organized as follows. In section 2, we
present results for the EW synchronization problem on weighted
uncorrelated SF networks from a constrained optimization viewpoint.
In section~3, we discuss the related questions for idealized transport
problems: weighted resistor networks and weighted random walks. A brief
summary is given in section~4.

\section{Optimization of Synchronization in Weighted Complex Networks in Noisy Environments}

The EW process on a network is given by the Langevin equation
\begin{equation}
\partial_{t} h_i = - \sum_{j=1}^{N} C_{ij}(h_i-h_j) +
\eta_{i}(t)\;,
\label{EW_ntwk}
\end{equation}
where $h_{i}(t)$ is the general stochastic field variable on a node
(such as fluctuations in the task-completion landscape in certain
distributed parallel schemes on computer networks
\cite{KORNISS03,KHK04,KHK05})
and $\eta_{i}(t)$ is a delta-correlated noise with zero mean and variance
$\langle\eta_{i}(t)\eta_{j}(t')\rangle$$=$$2\delta_{ij}\delta(t-t')$.
Here, $C_{ij}$$=$$C_{ji}$$>$$0$ is the symmetric coupling strength between the nodes
$i$ and $j$ ($C_{ii}$$\equiv$$0$). Defining the network Laplacian,
\begin{equation}
\Gamma_{ij}\equiv\delta_{ij}C_{i}-C_{ij}\;,
\label{laplacian_ntwk}
\end{equation}
where
$C_{i}\equiv\sum_{l}C_{il}$, we can rewrite Eq.~(\ref{EW_ntwk})
\begin{equation}
\partial_{t} h_i = - \sum_{j=1}^{N} \Gamma_{ij} h_j +
\eta_{i}(t)\;.
\label{EW_ntwk_gamma}
\end{equation}
For the steady-state equal-time two-point correlation function one finds
\begin{equation}
G_{ij} \equiv \langle(h_{i}-\bar{h})(h_{j}-\bar{h})\rangle =
\hat{\Gamma}^{-1}_{ij} =
\sum_{k=1}^{N-1} \frac{1}{\lambda_{k}}\psi_{ki}\psi_{kj} \;,
\label{corr_func}
\end{equation}
where $\bar{h}=(1/N)\sum_{i=1}^{N}h_i$ and
$\langle\ldots\rangle$ denotes an ensemble average over the noise
in Eq.~(\ref{EW_ntwk_gamma}). Here, $\hat{\Gamma}^{-1}$ denotes the inverse
of $\Gamma$ in the space orthogonal to the zero mode. Also, $\{\psi_{ki}\}_{i=1}^{N}$ and
$\lambda_{k}$, $k=0,1,\dots,N-1$, denote the $k$th normalized eigenvectors and the corresponding
eigenvalues, respectively. The $k=0$ index is reserved for the zero mode of the
Laplacian on the network: all components  of this eigenvector are identical and $\lambda_{0}=0$.
The last form in Eq.~(\ref{corr_func}) (the spectral decomposition of $\hat{\Gamma}^{-1}$) 
can be used to directly employ the results of exact numerical diagonalization.
The average steady-state spread or width in the synchronization
landscape can be written as \cite{Korniss_PLA,KHK04,KHK05}
\begin{equation}
\langle w^2 \rangle \equiv
\left\langle\frac{1}{N}\sum_{i=1}^{N}(h_i-\bar{h})^2\right\rangle =
\frac{1}{N}\sum_{i=1}^{N} G_{ii} =
\frac{1}{N}\sum_{k=1}^{N-1} \frac{1}{\lambda_{k}}\;.
\label{w2_def}
\end{equation}
The above observable is typically self-averaging (confirmed by numerics), i.e., 
the width $\langle w^2 \rangle$ for a sufficiently large, single network realization  
approaches the width averaged over the network ensemble.

The focus of this section is to optimize synchronization (i.e., minimize the width) on
({\it i}) weighted uncorrelated networks with SF degree distribution, ({\it ii})
subject to  fixed a cost. In the context of this work, we define the
total cost $C_{tot}$ simply to equal to the sum of weights over all edges in
the network
\begin{equation}
\sum_{i<j}C_{ij} = \frac{1}{2}\sum_{i,j}C_{ij} = C_{tot}\;.
\label{cost}
\end{equation}
The elements of the coupling matrix $C_{ij}$ can be expressed in terms
of the network's adjacency matrix $A_{ij}$ and the assigned weights $W_{ij}$ 
connecting node $i$ and $j$ as
$C_{ij}=W_{ij}A_{ij}$.
Here, we consider networks where the weights are symmetric and
proportional to a power of the degrees of the two nodes connected by
the link, $W_{ij}\propto(k_ik_j)^{\beta}$. We choose our cost
constraint to be such that it is equal to that of the unweighted
network, where each link is of unit strength.
\begin{equation}
\sum_{i,j}C_{ij} = 2C_{tot} = \sum_{i,j}A_{ij}  = N\overline{k}\;,
\label{cost_spec}
\end{equation}
where $\overline{k}=\sum_{i}k_{i}/N=\sum_{i,j}A_{ij}/N$ is the mean degree of the graph, i.e., the
average cost per edge is fixed. Thus, the question we ask, is how to
allocate the strength of the links in networks with heterogeneous
degree distributions with fixed total cost in order to optimize
synchronization. That is, the task is to determine the value of
$\beta$ which minimizes the width Eq.~(\ref{w2_def}), subject to the
constraint Eq.~(\ref{cost_spec}).

Combining the form of the weights, $W_{ij}\propto(k_ik_j)^{\beta}$,
and the constraint Eq.~(\ref{cost_spec}) one can immediately
write for the coupling strength between nodes $i$ and $j$
\begin{equation}
C_{ij} = N\overline{k}\frac{A_{ij}(k_ik_j)^{\beta}}{\sum_{l,n}A_{ln}(k_l k_n)^{\beta}}
\label{C_ij}
\end{equation}
From the above it is clear that the distribution of the weights is
controlled by a single parameter $\beta$, while the total cost is
fixed, $C_{tot}=N\overline{k}/2$.

\subsection{The globally optimal network with fixed cost}

Before tackling the above optimization problem for the restricted set
of heterogeneous networks and the specific form of weights, one may ask what is
the optimum among all networks with fixed cost, for which the EW synchronization
problem yields the minimum width. This will serve as a ``baseline''
reference for our problem. From the above framework it follows that
\begin{equation}
2C_{tot} = \sum_{i,j}C_{ij} = \sum_{i}C_{i} = \sum_{i}\Gamma_{ii} = \Tr(\Gamma) =
\sum_{l\neq 0}\lambda_{l}\;.
\end{equation}
Thus, the global optimization problem can be cast as
\begin{equation}
\langle w^2 \rangle =
\frac{1}{N}\sum_{l=1}^{N-1} \frac{1}{\lambda_{l}} = \mbox{minimum}\;,
\label{w2_l}
\end{equation}
with the constraint
\begin{equation}
\sum_{l=1}^{N-1}\lambda_{l} = 2C_{tot} = \mbox{fixed}\;.
\label{constraint}
\end{equation}
This elementary extremum problem, Eqs.~(\ref{w2_l}) and
(\ref{constraint}), immediately yields a solution where all $N$$-$$1$ non-zero
eigenvalues are equal,
\begin{equation}
\lambda_{l} = \frac{2C_{tot}}{N-1} \;,\;\;\;\;\; l=1,2,\ldots,N-1\;,
\end{equation}
and the corresponding {\em absolute} minimum of the width is
\begin{equation}
\langle w^2 \rangle_{\rm min} = \frac{(N-1)^2}{2NC_{tot}}\;.
\end{equation}
As one can easily see, the above set of identical eigenvalues corresponds
to a coupling matrix and network structure where each node is connected to all others with
identical strength  $C_{ij}=2C_{tot}/[N(N-1)]$. That is, for fixed
cost, the {\em fully connected} (FC) network is optimal, yielding the absolute
minimum width.

If we now consider the synchronization problem on any
network with $N$ nodes, with average degree $\overline{k}$, with
total cost $C_{tot}=N\overline{k}/2$ to be optimized with respect to the single parameter $\beta$,
the above result provides an absolute lower bound for the optimal width
\begin{equation}
\langle w^2(\beta)\rangle_{\rm min}\geq
\frac{(N-1)^2}{N^{2}}\frac{1}{\overline{k}}\simeq
\frac{1}{\overline{k}}\;.
\label{global_w2min}
\end{equation}

While the above results for the FC network provide a {\em mathematical}
absolute upper bound for the synchronization efficiency (absolute
lower bound for the width), one may wonder about the technological
feasibility of FC networks from a system-design viewpoint.
While, clearly, the performance of the FC network in noisy synchronization
is theoretically optimal among all networks of the same cost, in many realistic scenarios,
it is not realizable in the large-$N$ limit: For fixed total
cost of $C_{tot}\sim{\cal O}(N)$ the link strength is of ${\cal
O}(1/N)$. In ``hard-wired'' infrastructure networks, there is a minimal cost
per link, i.e., one cannot construct links for an arbitrary low
infinitesimal [${\cal O}(1/N)$] cost. Thus, in actual applications it
may not be possible to trade sparse networks with ${\cal O}(1)$ average
degree with links of strength ${\cal O}(1)$ for
fully-connected ones with the same cost with an average degree ${\cal
O}(N)$ with links of strength ${\cal O}(1/N)$.

In abstract (``logical'') communication networks, where in principal, each node can
communicate with all others (but the actual messages are routed through
a ``physical'' sparse hard-wired network) a FC logical communication
network can be realized in a dynamic (or ``annealed'') fashion \cite{KHK05}. At each
time step, each node communicates with another one chosen at random with a given frequency.

.

\subsection{Mean-field approximation on uncorrelated SF
networks}

First, we approximate the equations of motion [Eq.~(\ref{EW_ntwk})] by
replacing the local weighted average field $(1/C_i)\sum_{j}C_{ij}h_j$
with the global average $\overline{h}$ (the mean--height)
\begin{eqnarray}
\fl \partial_{t} h_i = - \sum_{j=1}^{N} C_{ij}(h_i-h_j) + \eta_{i}(t)
= -C_{i}\left(h_i- \frac{\sum_{j}C_{ij}h_j}{C_i}\right) +
\eta_{i}(t) \nonumber \\
\approx  - C_{i}\left(h_i- \overline{h}\right) +
\eta_{i}(t) \;.
\end{eqnarray}
As can be directly seen by summing up Eq.~(\ref{EW_ntwk}) over all
nodes, the mean height $\overline{h}$ performs a simple random walk with
noise intensity ${\cal O}(1/N)$. Thus, in the mean-field (MF)
approximation, in the asymptotic large-$N$ limit,
fluctuations {\em about the mean} are decoupled and reach a stationary
distribution with variance
\begin{equation}
\left\langle(h_i-\bar{h})^2\right\rangle\approx1/C_i \;,
\label{h_i}
\end{equation}
yielding
\begin{equation}
\langle w^2 \rangle =
\frac{1}{N}\sum_{i=1}^{N}\left\langle(h_i-\bar{h})^2\right\rangle
\approx
\frac{1}{N}\sum_{i}\frac{1}{C_i}\;.
\label{w2_1}
\end{equation}

Next, we establish an approximate relationship between the effective
coupling to the mean, $C_i$, and the degree $k_i$ of node $i$, for
{\em uncorrelated} (UC) weighted random graphs. Using the specific form of the
weights as constructed in Eq.~(\ref{C_ij}), we write
\begin{equation}
C_i =\sum_{j}C_{ij} 
= N\overline{k}\frac{\sum_{j}A_{ij}(k_ik_j)^{\beta}}{\sum_{l,n}A_{ln}(k_l k_n)^{\beta}}
= N\overline{k}\frac{k_i^{\beta}\sum_{j}A_{ij}k_j^{\beta}}{\sum_{l}k_l^{\beta}\sum_{n}A_{ln}k_n^{\beta}}\;.
\label{C1}
\end{equation}
For large minimum (and in turn, average) degree, expressions of the
form $\sum_{j}A_{ij}k_j^{\beta}$ can be approximated as
\begin{eqnarray}
\fl \sum_{j}A_{ij}k_j^{\beta} = \left(\sum_{j}A_{ij}\right)
\frac{\sum_{j}A_{ij}k_j^{\beta}}{\sum_{j}A_{ij}} 
= k_i \frac{\sum_{j}A_{ij}k_j^{\beta}}{\sum_{j}A_{ij}} \nonumber \\
\approx k_i \int dk P(k|k_i)k^{\beta}\;,
\end{eqnarray}
where $P(k|k')$ is the probability that an edge
from node with degree $k'$ connects to a node with degree $k$. For
{\em uncorrelated} random graphs, $P(k|k')$ does {\em not} depend on $k'$,
one has $P(k|k')=kP(k)/\langle k\rangle$ \cite{MendesREV,Vespignani_book}, where
$P(k)$ is the degree distribution and $\langle k\rangle$ is the
ensemble-averaged degree. Thus, Eq.~(\ref{C1}), for UC random networks, can be approximated as
\begin{eqnarray}
\fl C_i 
\approx N\langle k\rangle \frac{k_i^{\beta+1}\int dk P(k|k_i)k^{\beta}}
{N \int dk' k'^{\beta+1}P(k')  \int dk P(k|k')k^{\beta}} \nonumber \\
= \langle k\rangle \frac{k_i^{\beta+1}}{\int_{m}^{\infty} dk' k'^{\beta+1}P(k')}\;.
\label{C2}
\end{eqnarray}
Here, we consider SF degree distributions,
\begin{equation}
P(k)=(\gamma-1)m^{\gamma-1}k^{-\gamma}\;,
\label{P_k}
\end{equation}
where $m$ is the minimum degree in the network and $2<\gamma\leq3$. The average and the
minimum degree are related through $\langle k\rangle=m(\gamma-1)/(\gamma-2)$. No upper cutoff is
needed for the convergence of the integral in Eq.~(\ref{C2}), provided
that $2+\beta-\gamma<0$, and one finds
\begin{equation}
C_i \approx
\frac{\gamma-2-\beta}{\gamma-2}\frac{k_i^{\beta+1}}{m^{\beta}} \;.
\label{C3}
\end{equation}
Thus, for uncorrelated random SF graphs with large minimum
degree, the effective coupling coefficient $C_i$
only depends on the degree $k_i$ of node $i$, i.e., for a node with degree
$k$
\begin{equation}
C(k) \approx
\frac{\gamma-2-\beta}{\gamma-2}\frac{k^{\beta+1}}{m^{\beta}} \;.
\label{Ck}
\end{equation}
Finally, assuming self-averaging for large enough networks and combining the above, one obtains
for the width of the synchronization landscape
\begin{eqnarray}
\fl \langle w^2(\beta)\rangle 
\approx \frac{1}{N}\sum_{i}\frac{1}{C_i} \approx \int_{m}^{\infty} dk P(k)\frac{1}{C(k)}
\nonumber \\
= \frac{1}{\langle k\rangle} \frac{(\gamma-1)^2}{(\gamma-2-\beta)(\gamma+\beta)}\;,
\label{w2_2}
\end{eqnarray}
where using infinity as the upper limit is justified for $\gamma+\beta>0$.
Elementary analysis yields the main features of the above expression
for the average width:
\begin{enumerate}
\item $\langle w^2(\beta)\rangle$ is minimum at $\beta=\beta^{*}=-1$,
{\em independent} of the value of $\gamma$.
\item $\langle w^2\rangle_{\rm min} = \langle w^2(\beta^{*})\rangle = 1/\langle k\rangle$
\end{enumerate}
The above approximate result is consistent with using infinity as the
upper limit in all integrals, in that the optimal value $\beta^{*}=-1$
falls inside the interval $-\gamma<\beta<\gamma-2$ for
$2<\gamma\leq3$.
Interestingly, one can also observe, that, in this approximation, the
minimal value of the width is equal to that of the global optimum
[Eq.~(\ref{global_w2min})], realized by the fully connected network of
the same cost $N\langle k\rangle/2$, i.e. with identical links of strength
$\langle k\rangle/(N-1)$.

We emphasize that in obtaining the above result [Eq.~(\ref{w2_2})] we
employed two very strong and distinct assumptions/approximations:
({\it i}) for the dynamics on the network, we neglected
correlations (in a MF fashion) between the local field
variables and approximated the local height fluctuations by Eq.~(\ref{h_i});
({\it ii}) we assumed that the
network has no degree-degree correlations between nodes which are
connected (UC), so that the ``weighted degree'' $C_i$ can be approximated
with Eq.~(\ref{C3}) for networks with $m$$\gg$$1$.

\subsection{Numerical results}

For comparison with the above mean-field results, we considered
Barab\'asi-Albert (BA) SF networks \cite{Barab_sci,BarabREV}
\footnote{For the BA scale-free model \cite{Barab_sci} (growth and preferential
attachment), each new node is connected to the network with $m$ links,
resulting in $\langle$k$\rangle\simeq2m$ in the large-$N$ limit. Here, we employed a 
fully-connected initial cluster of $m+1$ nodes.}, 
``grown'' to $N$ nodes, where $P(k)=2m^2/k^3$, i.e., $\gamma=3$. 
While growing
networks, in general, are not uncorrelated, degree-degree correlations
are anomalously (marginally) weak for the  BA network
\cite{Vespignani_book,Catanzaro_PRE05}.
\begin{figure}[t]
\centering
\vspace*{1.80truecm}
       \includegraphics{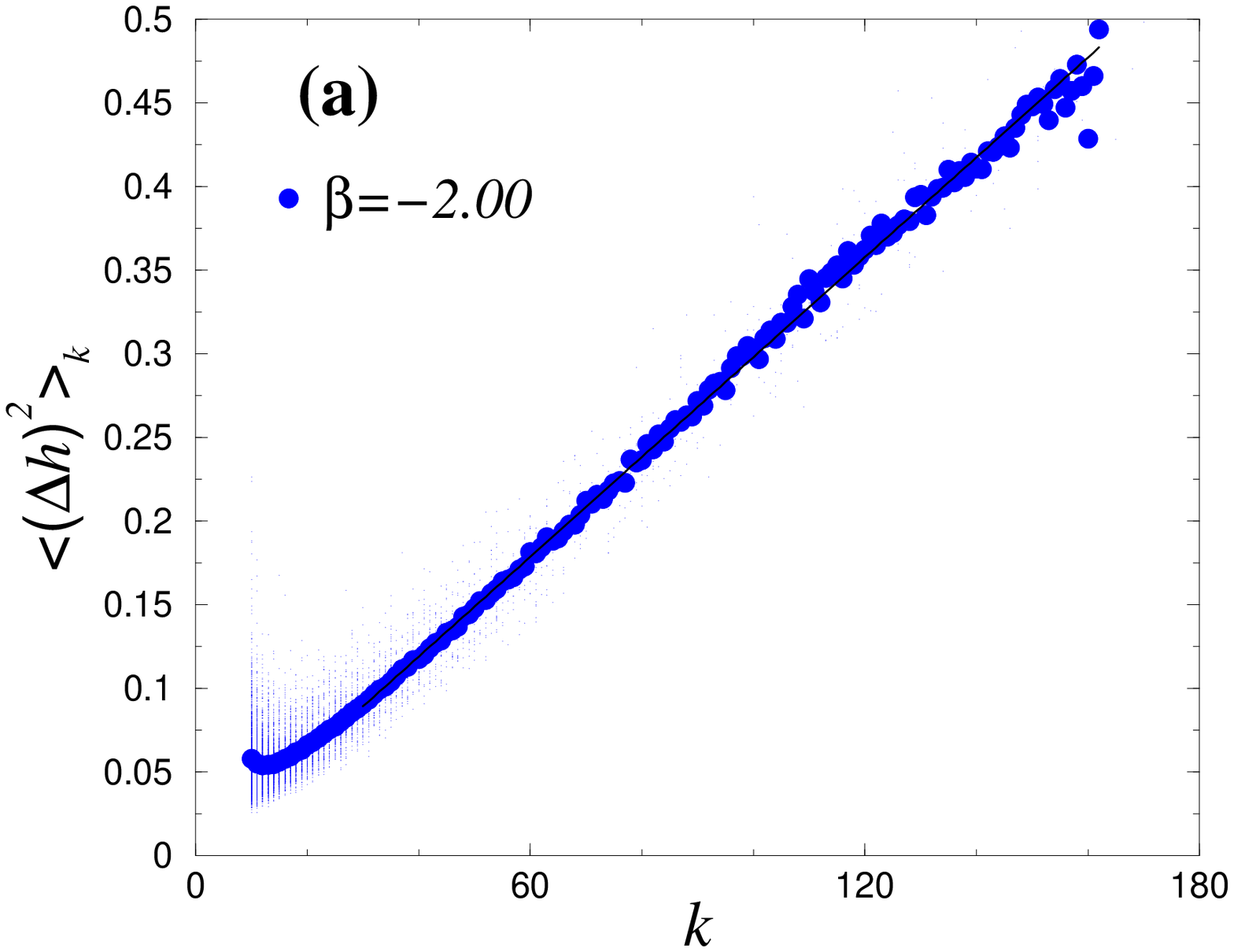}
       \includegraphics{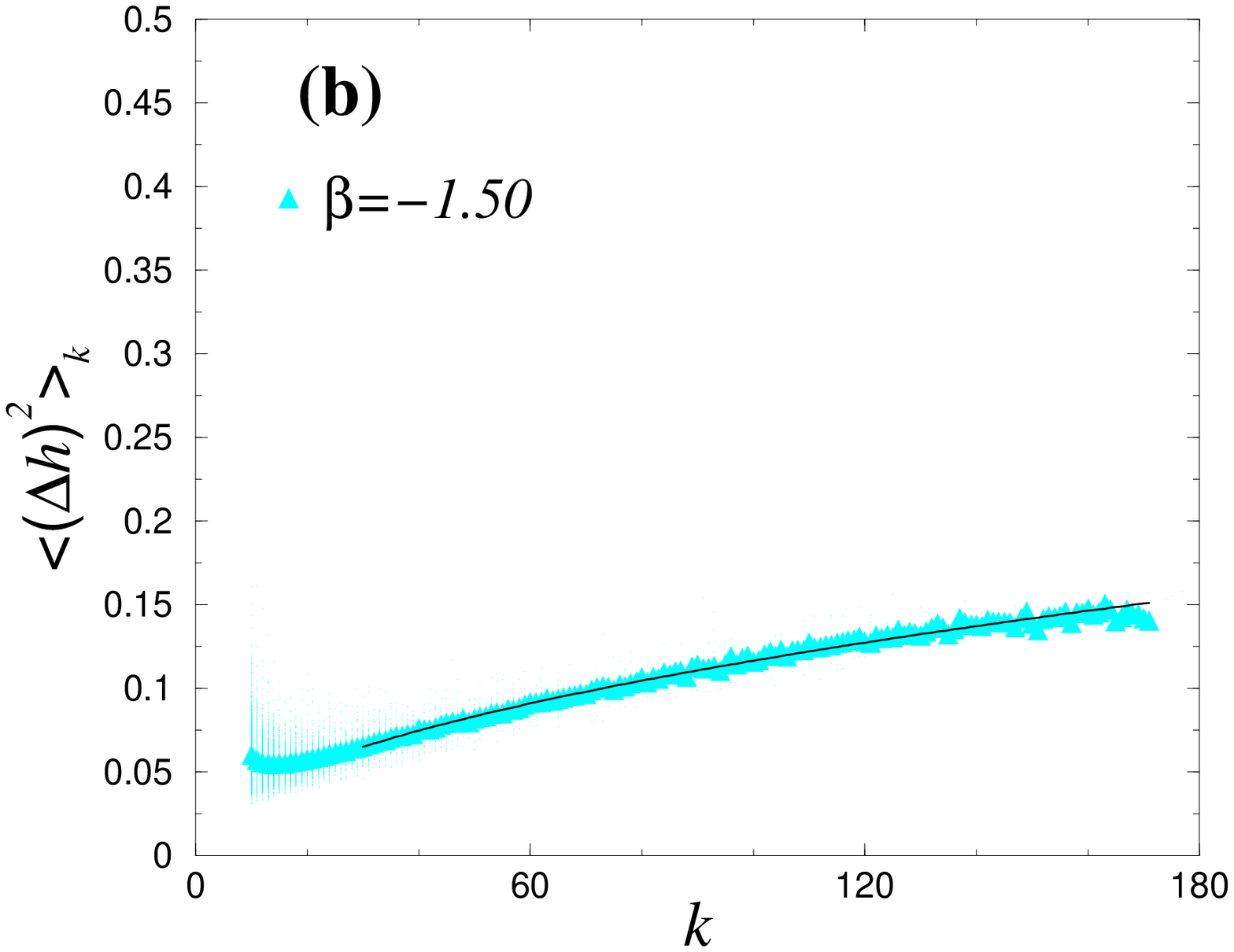}
\vspace*{5.00truecm}
       \includegraphics{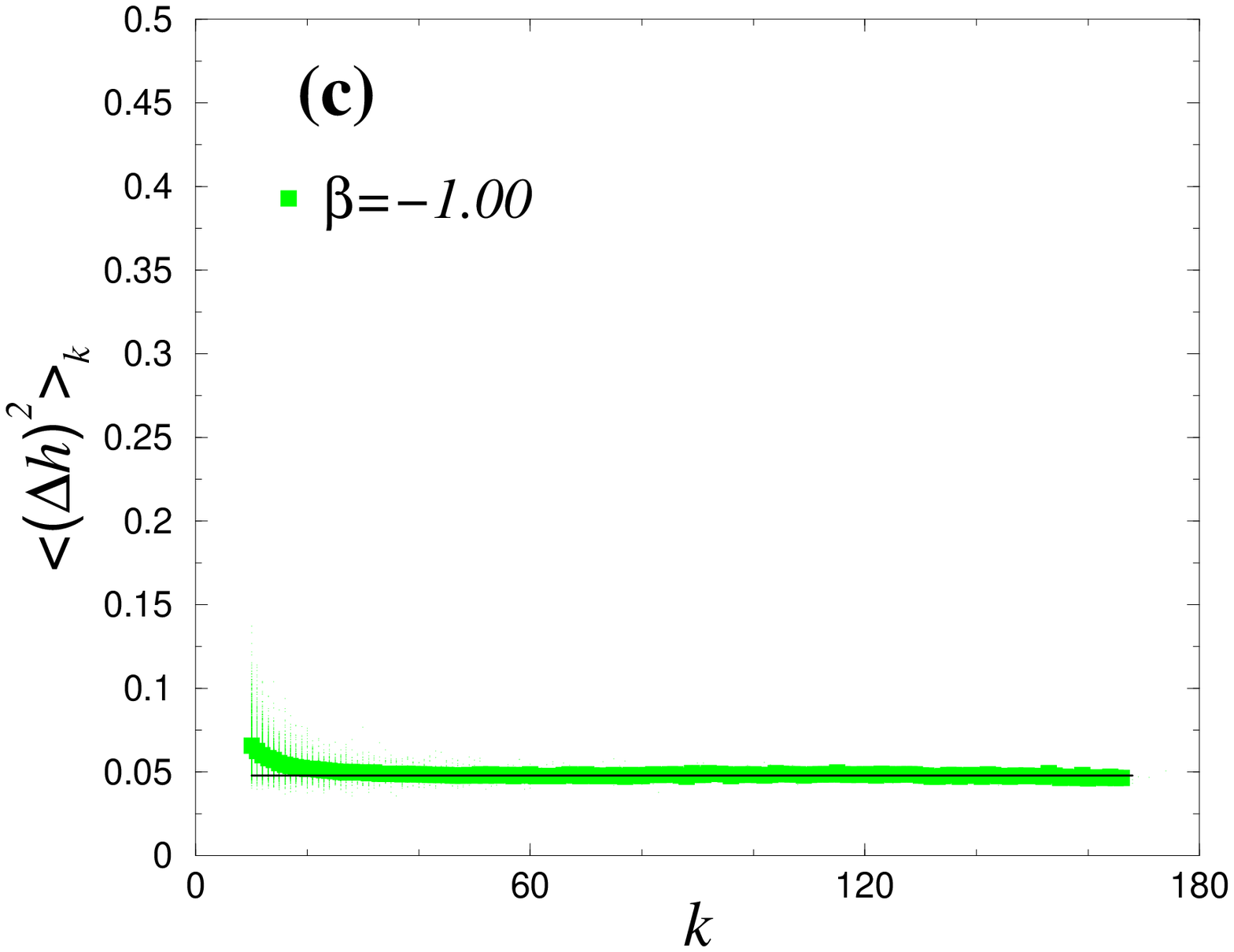}
       \includegraphics{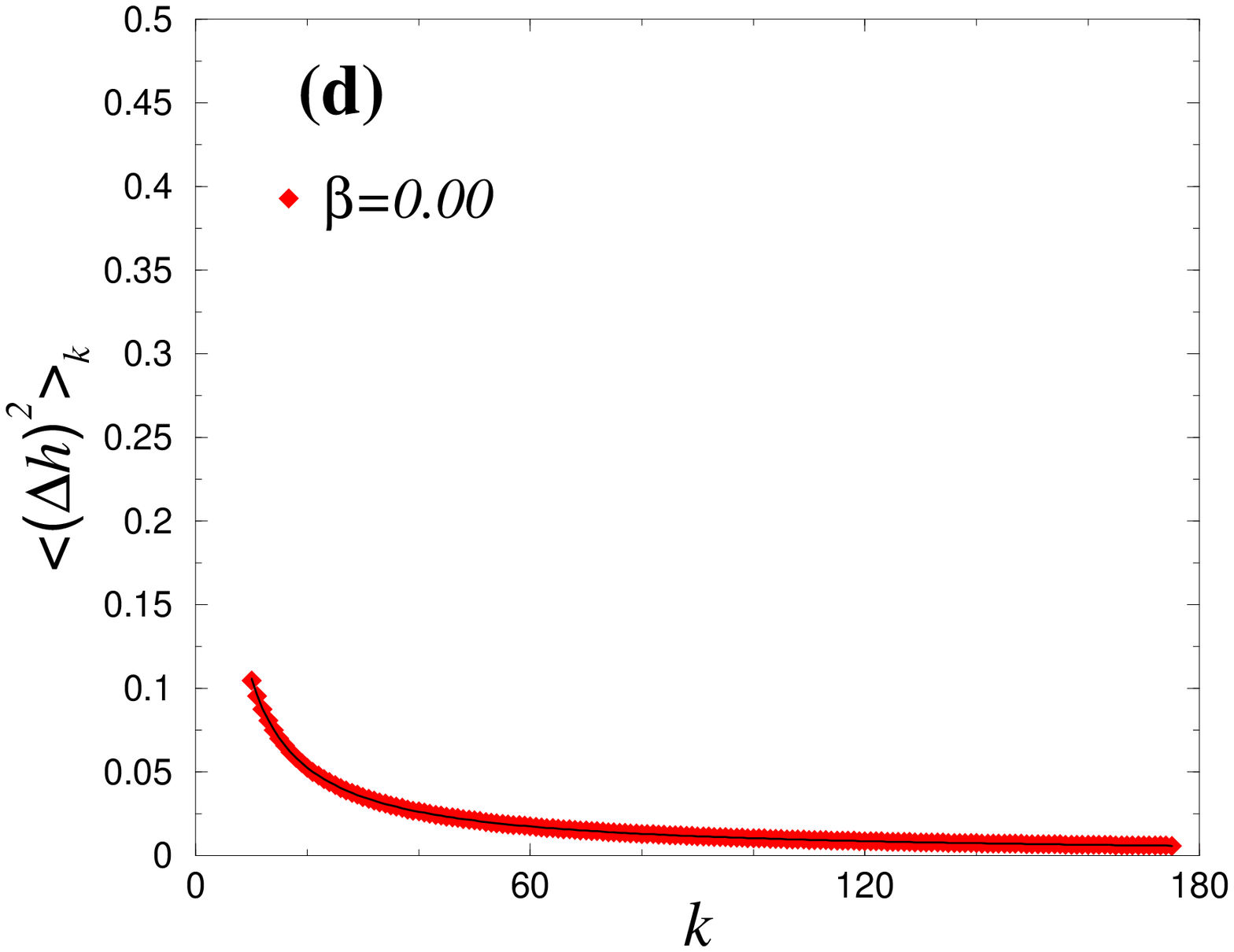}
\vspace*{3.50truecm}
\caption{Height fluctuations as a function of the degree of
       the nodes for $N$$=$$1000$, $\langle$$k$$\rangle=20$, and
       for (a) $\beta$$=$$-2.00$, (b) $\beta$$=$$-1.50$, (c)
       $\beta$$=$$-1.00$, and (d) $\beta$$=$$0.00$.
       Data, represented by filled symbols, are averaged over all nodes with
       degree $k$. Scatter plot (dots) for individual nodes is also shown from ten
       network realizations. Solid lines correspond to the MF+UC
       scaling $\langle(\Delta h)^2\rangle_{k}\sim k^{-(\beta+1)}$.}
\label{fig1}
\end{figure}

We have performed exact numerical diagonalization and employed
Eq.~(\ref{corr_func}) to find the local height fluctuations and
Eq.~(\ref{w2_def}) to obtain the width for a given network realization.
We carried out the above procedure for $10$--$100$ independent network
realizations. Finite-size effects (except for the $m$$=$$1$ BA tree network) are very weak for
$-2<\beta<1$; the width essentially becomes independent of the system
size. Figure~\ref{fig1} displays result for the local height fluctuations as a function of the
degree of the node. We show both the fluctuations averaged over all
nodes with degree $k$ and the scattered data for individual nodes. One
can observe that our approximate results for the scaling with the
degree [combining Eqs.~(\ref{h_i}) and (\ref{C3})],
$\left\langle(h_i-\bar{h})^2\right\rangle\approx1/C_i\sim k_{i}^{-(\beta+1)}$,
work very
well, except for very low degrees. The special case $\beta$$=$$0$, is
exceptionally good, since here $C_i=\sum_{j}A_{ij}=k_i$ exactly, and
the only approximation is Eq.~(\ref{h_i}).

\begin{figure}[t]
\centering
\vspace*{2.3truecm}
       \includegraphics{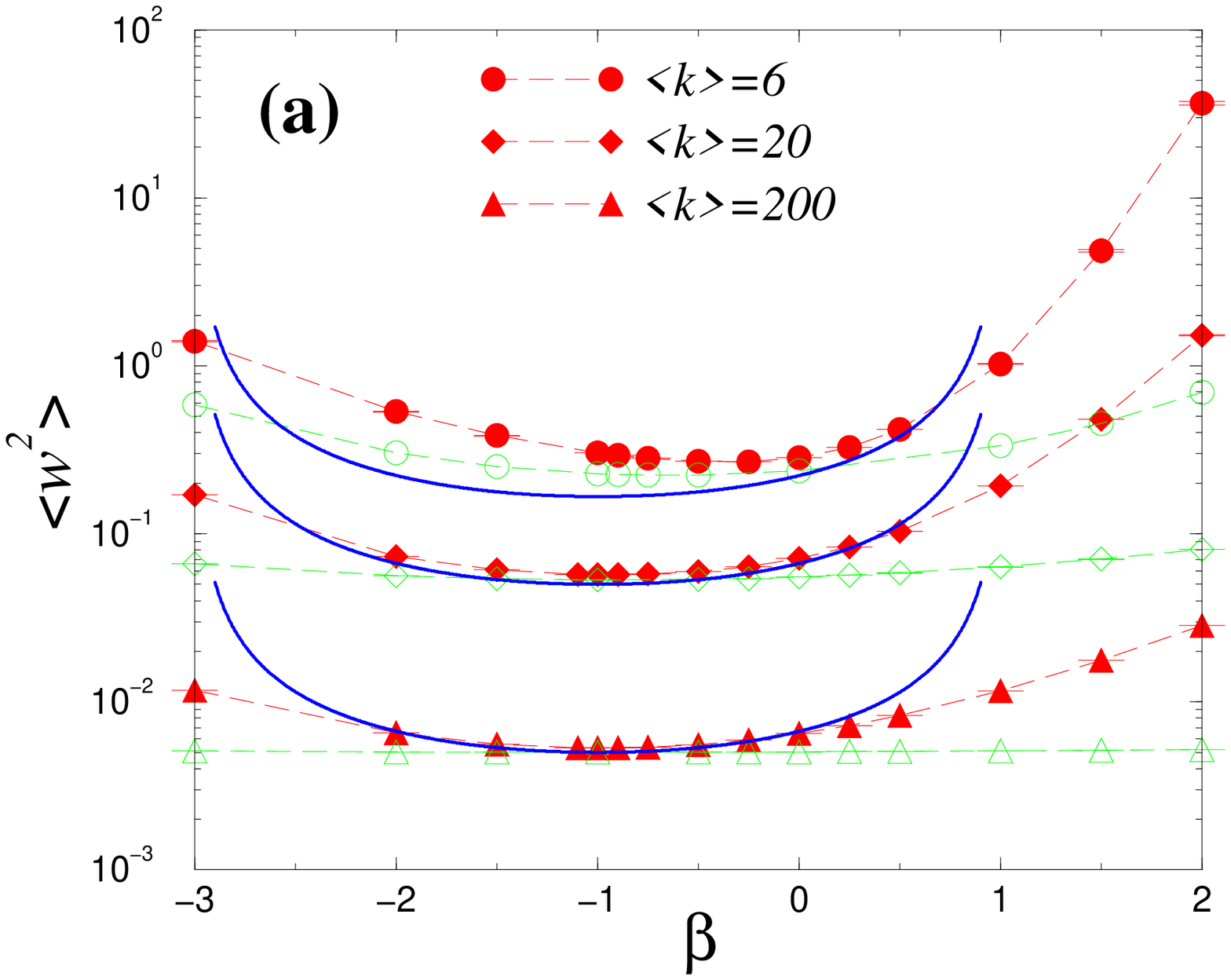}
       \includegraphics{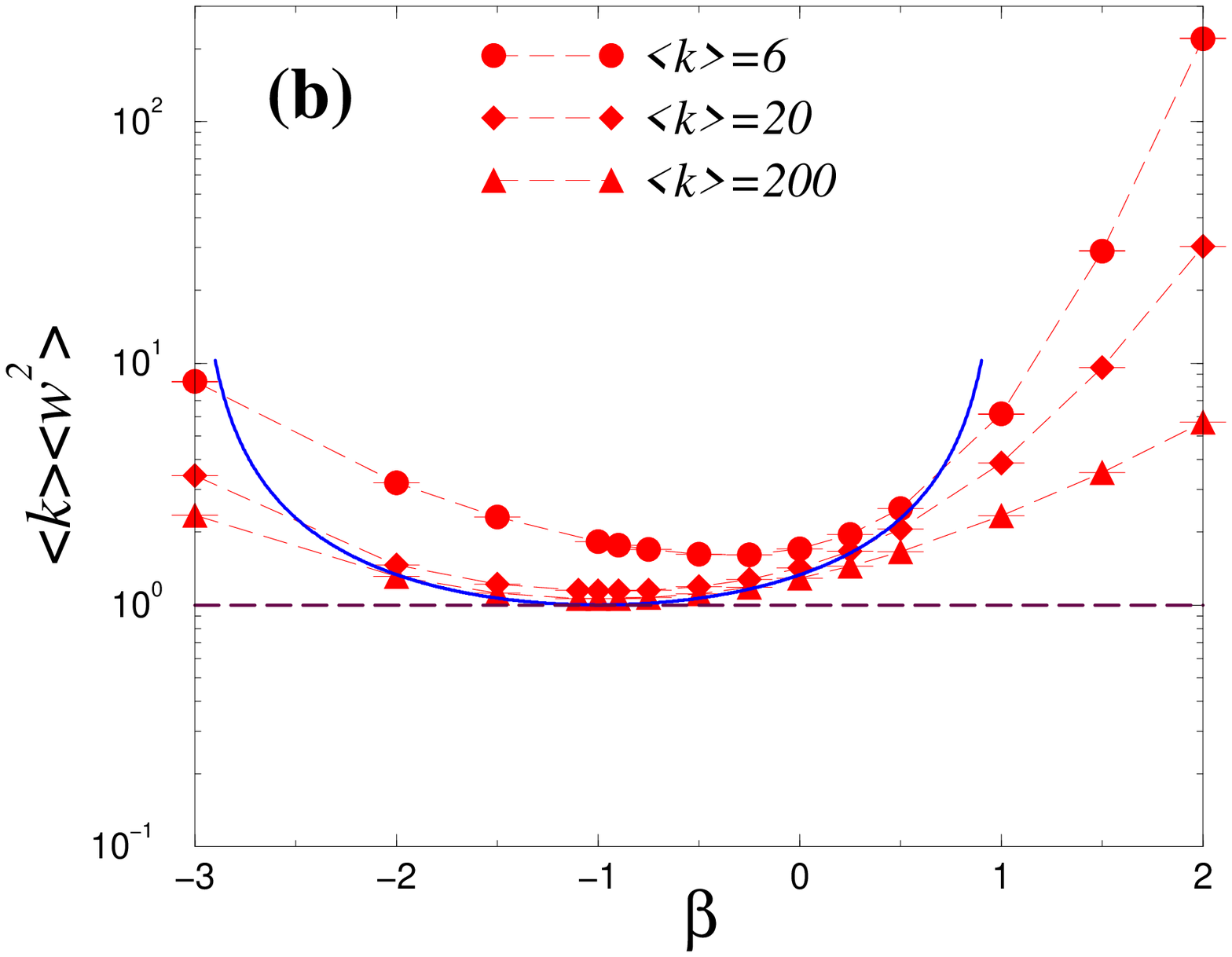}
\vspace*{3.50truecm}
\caption{(a) Steady-state width of the EW synchronization landscape as
       a function of the weighting parameter
       $\beta$ for the BA networks for $N$$=$$1,000$ with various average degree
       $\overline{k}\simeq\langle k\rangle\simeq 2m$.  Solid curves are the scaled approximate (MF+UC)
       results [Eq.~(\ref{w2_2})] for the same degree.
       For comparison, numerical results for SW networks with
       the same degree (with the respective open symbols) are also
       shown. Also, see Table~\ref{table1} for
       actual numerical values for $\beta$$=$$-1$.
       (b) Scaled width as a function of the weighting parameter
       $\beta$. The solid curve is the scaled approximate (MF+UC)
       result [Eq.~(\ref{w2_2})]. The horizontal dashed line indicates the (similarly
       scaled) absolute lower bound, as achieved by the fully
       connected network with the same cost $N\langle k\rangle/2$.}
\label{fig2}
\end{figure}

In Fig.~\ref{fig2}, we show our numerical results for the width and
compare it with the approximate (MF+UC) results Eq.~(\ref{w2_2}). They
agree reasonably well for networks with $m\gg1$. The divergence of
the approximate result Eq.~(\ref{w2_2}) at $\beta$$=-3$ and $\beta$$=1$
is the artifact of using infinity as the upper limit in the integrals performed
in our approximations.

\begin{figure}[t]
\centering
\vspace*{3.30truecm}
       \includegraphics{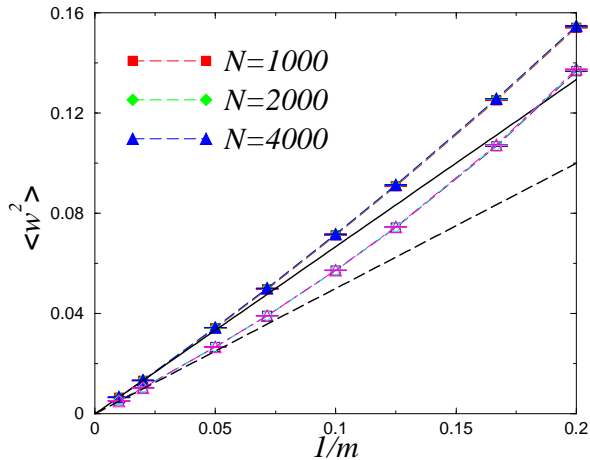}
\vspace*{3.50truecm}
\caption{Steady-state width of the EW synchronization landscape
       as a function of $1/m$ for the BA networks for
       $\beta=0.00$ (solid symbols) and $\beta=-1.00$ (respective open
       symbols), for three different system sizes.
       Straight lines (solid for $\beta$$=$$0$ and dashed for $\beta$$=$$-1$) 
       correspond to the MF+UC approximation Eq.~(\ref{w2_2}).}
\label{fig3}
\end{figure}

The results for the width clearly indicate the existence of a minimum
at a value of $\beta^*$ somewhat greater than $-1$. As the minimum
degree $m$ is increased, the optimal $\beta$ approaches $-1$
from above. This is not surprising, since in the limit of $m\gg1$
(large minimum degree), both the MF and the UC part of our
approximations are expected to work progressively better. In
Fig.~\ref{fig3}, we show the width as a function of $1/m$ for the BA
networks, indicating the rate of convergence to the MF+UC result, Eq.~(\ref{w2_2}).
Fig.~\ref{fig3} also indicates that finite-size effects are very small
and only contribute as small corrections to the {\em finite} value of the width
in the linmit of $N$$\to$$\infty$.
For $\beta$$=$$0$, our approximation [Eq.~(\ref{w2_2})] is within
$8\%$, $4\%$, and $1\%$ of the numerical results (based
on the above exact numerical diagonalization procedure) for $m$$=$$10$,
$m$$=$$20$, and $m$$=$$100$, respectively.
For $\beta$$=$$-1$, it is within $15\%$, $7\%$, and $3\%$ of the numerical results for $m$$=$$10$,
$m$$=$$20$, and $m$$=$$100$, respectively. Thus, our
approximation works very well for large uncorrelated sparse SF
networks with sufficiently large, {\em finite} minimum (and consequently, average) degree.

The above optimal link-strength allocation at around the value
$\beta^*$$=$$-1$ seems to be present in all random networks where the
degree distribution is different from a delta-function.
For example, in SW networks, although the degree distribution has an exponential tail,
$\langle w^2\rangle$ also exhibits a minimum, but the effect is much
weaker, as shown in Fig.~\ref{fig2}(a).
Further, a point worthwhile to mention, a SW network with the same number of
nodes and the same average degree (corresponding to the same cost)
always ``outperforms'' its SF counterpart (in terms of minimizing the width). 
The difference between their
performance is smallest around the optimal value, where both are very
close to that of the lowest possible value, realized by the FC network of the same cost
(Table~\ref{table1}.)

\begin{table}
\caption{\label{table1}Comparing numerical values of the steady-state width $\langle w^2\rangle$ of the
EW process at $\beta$$=$$-1$ for BA and SW networks of the same
finite average degree $\langle k\rangle$ and cost $N\langle k\rangle/2$ (for
$N$$=$$1000$) with results of the MF+UC approximation Eq.~(\ref{w2_2}).
Note that the width in the MF+UC approximation for $\beta$$=$$-1$
coincides with that of the globally optimal FC network of the same cost
[compare Eqs.~(\ref{global_w2min}) and (\ref{w2_2})],
$\langle w^2\rangle\simeq 1/\langle k\rangle$.
Error bars on the numerically obtained values for
the BA and SW networks are less that the last digit shown in the table.}
\begin{indented}
\item[]\begin{tabular}{@{}llll}
  \br
     $\langle k\rangle$ & BA & SW & FC \\
  \mr
     6 &  0.304   & 0.228   &  0.1666 \\
     20 & 0.0571  & 0.0531  &  0.0500 \\
     200 & 0.0053 & 0.00501 &  0.0050 \\
  \br
\end{tabular}
\end{indented}
\end{table}

\section{Connections with Transport and Flow Problems in Weighted Networks}

\subsection{Optimizing the system resistance in weighted
resistor networks}

Resistor networks have been widely studied since the 70's as models
for conductivity problems and classical transport in disordered
media \cite{Kirkpatrick71,Kirkpatrick73}.
Amidst the emerging research on complex networks, resistor networks
have been employed to study and explore community structures in
social networks \cite{NEWMAN04,HUBER04,NEWMAN05}. Most recently, they
were utilized to study transport efficiency in SF
\cite{Andrade_PRL2005,Lopez2005} and SW networks \cite{Korniss_PLA}.
The work by L\'opez et al. \cite{Lopez2005} revealed
that in SF networks  \cite{Barab_sci,BarabREV} anomalous transport
properties can emerge, displayed by the power-law tail of distribution
of the network conductance. Now, we consider weighted resistor networks
subject to a fixed total cost (the cost of
each link is associated with its conductance).

In a recent paper we have shown that observables in the EW
synchronization problem and in (Ohmic) resistor networks are inherently
related through the spectrum of the network Laplacian
\cite{Korniss_PLA}. Consider an arbitrary (connected) network where
$C_{ij}$ is the conductance of the link between node $i$ and
$j$, with a current $I$ entering (leaving) the network at node $s$ ($t$).
Introducing the voltages measured from the mean at each node, $\hat{V}_{i} =
V_{i}-\bar{V}$, where $\bar{V}$$=$$(1/N)\sum_{i=1}^{N}V_i$, one
obtains \cite{Korniss_PLA}
\begin{equation}
\hat{V}_{i}= I(G_{is} - G_{it})\;.
\label{voltage_solution}
\end{equation}
Here, $G$ is the same network propagator discussed in the context of
the EW process, i.e. the inverse [Eq.~(\ref{corr_func})] of the network Laplacian
[Eq.~(\ref{laplacian_ntwk})] in the space orthogonal to the zero mode.
Applying Eq.~(\ref{voltage_solution}) to nodes $s$ and $t$, where
the voltage drop between these nodes is $V_{st}=\hat{V}_{s}-\hat{V}_{t}$,
one immediately obtains the
effective two-point resistance of the network between nodes  $s$ and
$t$ \cite{Korniss_PLA,WU2004},
\begin{equation}
R_{st} \equiv \frac{V_{st}}{I} = G_{ss} + G_{tt} -2G_{st} =
\sum_{k=1}^{N-1} \frac{1}{\lambda_{k}}(\psi_{ks}^2 + \psi_{kt}^2 - 2 \psi_{ks}\psi_{kt}) \,.
\label{R_solution}
\end{equation}
The spectral decomposition in Eq.~(\ref{R_solution}) is, again, useful to employ 
the results of exact numerical diagonalization. Comparing Eqs.~(\ref{corr_func})
and (\ref{R_solution}), one can see that
the two-point resistance of a network between node $s$ and $t$ is the
same as the steady-state {\em height-difference} correlation function of the EW process
on the network \cite{Korniss_PLA},
\begin{equation}
\langle (h_s-h_t)^2\rangle =
\langle [(h_s-\overline{h})-(h_t-\overline{h})]^2\rangle =
G_{ss} + G_{tt} -2G_{st} = R_{st} \,.
\label{R_C}
\end{equation}
For example, using the above relationship and the employing the MF+UC
approximation
\footnote{In the context of resistor networks, while there are no ``fields'', we
carry over the terminology ``mean-field'' (MF) from the associated EW synhronization problem. In
terms of the network propagator, the assumptions of the MF approximation can be
summarized as $G_{st}$$\ll$$G_{ss}$ for all $s$$\neq$$t$, and
$G_{ss}$$\simeq$$1/C_{s}$.},
one can immediately obtain the scaling of the typical value of the effective two-point 
resistance in weighted resistance networks, between two nodes with degrees $k_s$ and $k_t$,
\begin{equation}
R_{st} \simeq G_{ss} + G_{tt} \sim [k_{s}^{-(1+\beta)} + k_{t}^{-(1+\beta)}]=
\frac{k_{s}^{1+\beta} + k_{t}^{1+\beta}}{(k_{s}k_{t})^{1+\beta}}\;.
\label{R_st}
\end{equation}

A global observable, measuring transport efficiency, analogous to
the width of the synchronization landscape, is the average
two-point resistance \cite{Korniss_PLA,Lopez2005} (averaged over all
pairs of nodes, for a given network realization). Using Eq.~(\ref{R_C}) and exploiting the basic
properties of the Green's function, one finds
\begin{eqnarray}
\fl \bar{R} \equiv \frac{2}{N(N-1)}\sum_{s<t}R_{st} =
\frac{1}{N(N-1)}\sum_{s\neq t}R_{st} \nonumber \\
=  \frac{N}{N-1}2\langle w^2 \rangle \simeq 2\langle w^2 \rangle \;,
\label{R_w2}
\end{eqnarray}
i.e., in the asymptotic large system-size limit the average system
resistance of a given network is twice the steady-state width of the
EW process on the same network. Note that the above relationships,
Eqs.~(\ref{R_C}) and (\ref{R_w2}), are exact and valid for any graph.

The corresponding optimization problem for resistor networks then reads as follows:
For a fixed total cost, $C_{\rm tot}=\sum_{i<j}C_{ij}=N\langle k\rangle/2$,
where the link conductances are weighted according to Eq.~(\ref{C_ij}),
what is the value of $\beta$ which minimizes the average system
resistance $\overline{R}(\beta)$? Based on the above relationship between the
average system resistance and the steady-state width of the EW process
on the same graph [Eq.~(\ref{R_w2})], the answer is the same as was
discussed in section~2 [Eq.~(\ref{w2_2})]: $\beta^{*}$$=$$-1$ and
$\overline{R}_{\rm min}=2N/[(N-1)\langle k\rangle]\simeq2/\langle k\rangle$ in the mean-field
approximation on uncorrelated random SF networks. Numerical
results for $\overline{R}(\beta)$ are also provided for ``free''
by virtue of the connection Eq.~(\ref{R_w2}), once we have the
results for $\langle w^2(\beta)\rangle$.

\subsection{Connection with random walks on weighted networks and
congestion-aware local routing schemes}

Consider the weights $\{C_{ij}\}$ employed in the previous sections
and define a random walk (RW) with the transition probabilities \cite{Doyle}
\begin{equation}
P_{ij} \equiv \frac{C_{ij}}{C_{i}}\;
\label{P_ij}
\end{equation}
and recall that $C_i=\sum_{l}C_{il}$. $P_{ij}$ is the probability that
the walker currently at node $i$ will hop to node $j$ in the next
step. Note that because of the construction of the transition
probabilities (being a ratio), the issue of cost constraint
disappears from the problem. That is, any normalization prefactor associated
with the conserved cost [as in Eq.~(\ref{C_ij})] cancels out, and the only
relevant information is  $C_{ij}\propto A_{ij}(k_ik_j)^{\beta}$,
yielding
\begin{equation}
P_{ij} = \frac{C_{ij}}{C_{i}} =
\frac{A_{ij}(k_ik_j)^{\beta}}{\sum_{l}A_{il}(k_ik_l)^{\beta}} =
\frac{A_{ij}k_j^{\beta}}{\sum_{l}A_{il}k_l^{\beta}}\;.
\label{P_ij_spec}
\end{equation}
Conversely, the results are invariant for any
normalization/constraint, so for convenience, one can use the
normalized form of the $C_{ij}$ coefficients as given in
Eq.~(\ref{C_ij}).

Having a random walker starting at an arbitrary source node $s$,
tasked to arrive at an arbitrary target node $t$,
the above weighted RW model can be associated with a simple {\em
local} routing or search scheme \cite{Guimera_PRL04} where
packets are independently forwarded to a nearest neighbor, chosen
according to the transition probabilities Eq.~(\ref{P_ij_spec}), until
the target is reached. These probabilities contain only limited local
information, namely the degree of all neighboring nodes.
By construction, the associated local (stochastic) routing problem (section 3.2.3) does not
concern link strength (bandwidth) limitations but rather the processing/queuing capabilities of the
nodes, so the cost constraint, associated with the links, disappears form the problem.

\subsubsection{Node betweenness for weighted RWs}

In network-based transport or flow problems, the appropriate betweenness measure
is defined to capture the amount of traffic or information passing through a node or a link, i.e., the
load of a node or a link
\cite{MendesREV,Goh_PRL2001,Goh_PRE2005,Noh_PRL2004,
Vespignani_book,NEWMAN05,Freeman_1977,Freeman_1979}.
Here, our observable of interest is the {\em node betweenness} $B_i$ for a given routing
scheme \cite{Guimera_PRL04} (here, purely local and characterized by a single parameter
$\beta$): {\em the expected number of visits} to node $i$ for a
random walker originating at node $s$ (the source) before reaching
node $t$ (the target) $E_{i}^{s,t}$, summed over all source-target pairs.
For a general RW, as was shown by Doyle and Snell \cite{Doyle},
$E_{i}^{s,t}$ can be obtained using the framework of the equivalent
resistor-network problem (discussed in section~3.1). More specifically,
\begin{equation}
E_{i}^{s,t} = C_i(V_i-V_t)\;,
\label{E_ist}
\end{equation}
while a {\em unit} current is injected (removed) at the source (target)
node. Utilizing again the network propagator and Eq.~(\ref{voltage_solution}), one obtains
\begin{equation}
E_{i}^{s,t} = C_i(V_i-V_t) = C_i(\hat{V}_i-\hat{V}_t) =
C_i(G_{is} - G_{it} - G_{ts} + G_{tt})\;.
\label{E_ist2}
\end{equation}
For the node betweenness, one then obtains
\begin{eqnarray}
\fl B_i = \sum_{s\neq t} E_{i}^{s,t} =
\frac{1}{2}\sum_{s\neq t} (E_{i}^{s,t} + E_{i}^{t,s})
= \frac{1}{2}\sum_{s\neq t} C_i(G_{ss} + G_{tt} - 2G_{ts}) \nonumber \\
= \frac{C_i}{2}\sum_{s\neq t} R_{st} = \frac{C_i}{2}N(N-1)\overline{R} \;.
\label{B_i}
\end{eqnarray}
Note that the above expression is valid for any graph and for an arbitrary
weighted RW defined by the transition probabilities Eq.~(\ref{P_ij}).
As can be seen from Eq.~(\ref{B_i}), the node betweenness is
proportional to the product of a local topological
measure, the weighted degree $C_i$, and a global flow measure, the
average system resistance $\overline{R}$.  As a
specific case, for the unweighted RW ($\beta$$=$$0$)
$C_i=\sum_{l}A_{il}=k_i$, thus, the node betweenness is exactly
proportional to the degree of the node, $B_i=k_iN(N-1)\overline{R}/2$.

Using our earlier approximations and results for uncorrelated SF
graphs Eq.~(\ref{C3}) and (\ref{w2_2}), and the relationship between
the width and the average system resistance Eq.~(\ref{R_w2}), for
weighted RW, controlled by the exponent $\beta$, we find
\begin{eqnarray}
\fl B_i(\beta) = \frac{C_i}{2}N(N-1)\overline{R} = C_iN^2\langle w^2\rangle \nonumber \\
\approx N^2\frac{\gamma-1}{\gamma+\beta} \frac{k_i^{1+\beta}}{m^{1+\beta}} \;.
\label{B_UCMF}
\end{eqnarray}

First, we consider the average ``load'' of the network
\begin{equation}
\overline{B}= \frac{1}{N}\sum_{i} B_i =
\frac{\sum_{i}C_i}{2}(N-1)\overline{R} \;.
\label{B_avg}
\end{equation}
Similar to Eq.~(\ref{B_i}), the above expression establishes an exact
relationship between the average node betweenness of an arbitrary RW
[given by Eq.~(\ref{P_ij})] and
the observables of the associated resistor network, the total edge
cost and the average system resistance. For example, for
the $\beta$$=$$0$ case, $\overline{B}=\overline{k}N(N-1)\overline{R}/2$.
As noted earlier, for calculation purposes one is free to consider the
set of $C_{ij}$ coefficients given by Eq.~(\ref{C_ij}), which also leads
us to the following statement:

\noindent
{\it For a RW defined by the transition probabilities Eq.~(\ref{P_ij}),
the average RW betweenness is minimal when the average system
resistance of the associated resistor network with fixed total edge
cost (and the width of the associated noisy synchronization network)
is minimal.}

Utilizing again our earlier approximations and results for uncorrelated SF
graphs and the relationship between the width and the average system
resistance, we find
\begin{eqnarray}
\fl \overline{B}(\beta) =  \frac{\sum_{i}C_i}{2}(N-1)\overline{R}
= \left(\sum_{i}C_i\right) N\langle w^2\rangle \nonumber \\
\approx N^2 \frac{(\gamma-1)^2}{(\gamma-2-\beta)(\gamma+\beta)}
\label{B_avg_UCMF}
\end{eqnarray}
The average node betweenness is minimal for $\beta=\beta^*=-1$, for all $\gamma$.

\subsubsection{Commute times and hitting times for weighted RWs}

The hitting time $\tau_{st}$ is the expected number of steps for
the random walker originating at node $s$ to reach node $t$ for the
first time. The commute time is the expected number of steps for a
``round trip'' between nodes $s$ and $t$, $\tau_{st}+\tau_{ts}$.
Relationships between the commute time and the effective two-point
resistance have been explored and discussed in detail in several works 
\cite{Lovasz,Chandra,Tetali}. In its most
general form, applicable to weighted networks, it was shown by Chandra
et al. \cite{Chandra} that
\begin{equation}
\tau_{st}+\tau_{ts} = \left(\sum_{i}C_i\right) R_{st} \;.
\label{t_c}
\end{equation}
For the average hitting (or first passage) time, averaged over all pairs
of nodes, one then obtains
\begin{eqnarray}
\fl \overline{\tau} \equiv \frac{1}{N(N-1)}\sum_{s\neq t} \tau_{s,t} =
\frac{1}{2N(N-1)}\sum_{s\neq t} (\tau_{s,t} + \tau_{t,s}) \nonumber \\
= \frac{\sum_{i}C_i}{2N(N-1)}\sum_{s\neq t} R_{st} =
\frac{\sum_{i}C_i}{2}\overline{R} \;.
\label{tau_avg}
\end{eqnarray}
Comparing Eq.~(\ref{B_avg}) and (\ref{tau_avg}),
the average hitting time (the average travel time for packets to reach
their destinations) then can be written as
$\overline{\tau}=\overline{B}/(N-1)$.
Note that this relationship is just a specific realization of Little's
law \cite{Little,Allen}, in the context of general communication networks,
stating that the average time needed for a packet to reach its destination is
proportional to the total load of the network. Thus, the average hitting time
and the average node betweenness (only differring by a factor of $N$-$1$)
are minimized {\em simultaneously} for the same graph (as a function
of $\beta$, in our specific problem).

\subsubsection{Network congestion due to queuing limitations}

Consider the simplest local ``routing'' problem \cite{Guimera_PRL04,Danila_PRE06} in which packets are
generated at {\em identical} rate $\phi$ at each node. Targets for each
newly generated packet are chosen at uniformly random from the
remaining $N-1$ nodes. Packets perform independent, weighted RWs,
using the transition probabilities Eq.~(\ref{P_ij}), until
they reach their targets.
Further, the queuing/processing capabilities of the nodes are limited
and are identical, e.g. (without loss of generality) each node can send out one packet per
unit time. From the above it follows that the network is
congestion-free as long as
\begin{equation}
\phi\frac{B_i}{N-1} <1\;,
\label{phi}
\end{equation}
for {\em every} node $i$
\cite{Krause_PA04,Guimera_PRL04,Zhao_05,Toro_PRL06,Danila_PRL06}. 
As the packet creation rate $\phi$
(network throughput per node) is increased, congestion emerges at a critical value $\phi_c$
when the inequality in Eq.~(\ref{phi}) is first violated. Up to that
point, the simple model of independent random walkers (discussed in the previous subsections), can
self-consistently describe the average load landscape in the network.
Clearly, network throughput is limited by the most congested node (the one with the maximum
betweenness), thus
\begin{equation}
\phi_c = \frac{N-1}{B_{\max}}\;,
\label{phi_c}
\end{equation}
a standard measure to characterize the efficiency of
communication networks
\cite{Krause_PA04,Guimera_PRL04,Zhao_05,Toro_PRL06,Danila_PRL06}.

To enhance or optimize network throughput (limited by the onset of
congestion at the nodes), one may scale up the processing capabilities
of the nodes \cite{Zhao_05}, optimize the underlying network topology
\cite{Guimera_PRL04}, or optimize routing by finding pathways which
minimize congestion \cite{Toro_PRL06,Danila_PRE06,Danila_PRL06}. 
The above RW routing, controlled by the weighting parameter $\beta$,
is an example for the latter, where the task is to maximize global
network throughput by localy directing traffic. In general, congestion
can also be strongly influenced by ``bandwith'' limitations
(or collisions of packets), which are related to the edge betweenness,
and not considered here.

For $\beta$$>$$-1$, within our approximations, nodes with high
betweenness coincide with nodes with high
degree. Further, for nodes with high degree, the mean-field approach
on uncorrelated SF graphs is expected to work reasonably well.
In this region, the scaling behavior $B_{\max}$ is related to
that of the highest degree $k_{\max}$ in the graph of {\em finite} size $N$.
The scaling of the maximum degree with the system size, however, even for idealized
SF network models, is very sensitive to the details of the
network construction. For example, in the region of our interest, $2<\gamma\leq3$, for the
standard configuration model (CM) \cite{Molloy}, the largest degree is governed by
the {\em natural cutoff}, $k_{\max}\simeq mN^{1/(\gamma-1)}$
\cite{MendesREV,Boguna}, but this network has correlations,
especially between nodes with larger degrees \cite{Catanzaro_PRE05,Boguna}. So one
may use the MF+UC approximation, but should expect stronger corrections.
On the other hand, in a recent construction for SF networks,
the uncorrelated configurational model (UCM) \cite{Catanzaro_PRE05},
the resulting network is genuinely uncorrelated, and the largest
degree is governed by the {\em structural cutoff}, $k_{\max}\simeq
(\langle k\rangle N)^{1/2}$ \cite{Catanzaro_PRE05,Boguna,Burda,Chung}. Combining these cutoff
behaviors with Eq.~(\ref{B_UCMF}), for the CM scale-free network model with
the natural cutoff one has
\begin{eqnarray}
B^{CM}_{\max}(\beta) \approx N^2\frac{\gamma-1}{\gamma+\beta}
\frac{k_{\max}^{1+\beta}}{m^{1+\beta}}
\simeq
\frac{\gamma-1}{\gamma+\beta} N^{\frac{2\gamma+\beta-1}{\gamma-1}}  \;,
\label{B_max_CM}
\end{eqnarray}
and
\begin{equation}
\phi^{CM}_c(\beta) = \frac{N-1}{B_{\max}} \approx
\frac{\gamma+\beta}{\gamma-1} N^{-\frac{\gamma+\beta}{\gamma-1}}
\sim {\cal O}\left( N^{-\frac{\gamma+\beta}{\gamma-1}}\right) \;.
\label{phi_c_CM}
\end{equation}
Similarly, for the UCM scale-free network model one finds
\begin{eqnarray}
B^{UCM}_{\max}(\beta)
\simeq
\frac{\gamma-1}{\gamma+\beta} N^2
\left(\frac{\gamma-1}{\gamma-2}\frac{N}{m}\right)^{\frac{1+\beta}{2}}  \;,
\label{B_max_UCM}
\end{eqnarray}
and
\begin{equation}
\phi^{UCM}_c(\beta) \simeq
\frac{\gamma+\beta}{\gamma-1}
\frac{1}{N}\left(\frac{\gamma-1}{\gamma-2}\frac{N}{m}\right)^{-\frac{1+\beta}{2}}
\sim {\cal O}\left( N^{-\frac{3+\beta}{2}}\right) \;.
\label{phi_c_UCM}
\end{equation}
From the above expression one can see that in the $\beta$$>$$-1$
region, for large $N$, the exponential decay in $\beta$ dominates
for both the CM [Eq.~(\ref{phi_c_CM})] and UCM [Eq.~(\ref{phi_c_UCM})]
scale-free networks. Consequently, in the semi-infinite region $\beta$$>$$-1$,
$\phi_c(\beta)$ is a monotonically decreasing function of $\beta$.

For $\beta$$<$$-1$, nodes with high betweenness are the nodes
with a low degree, but for these nodes the above approximations are
expected to work poorly. Further, there are many nodes with a degree of order $m$, and
the actual distribution of the betweenness [through the weighted degrees 
$C_i$, Eq.~(\ref{B_i})] for nodes with with $k_i$$\sim$$m$, depends strongly 
on the ``local'' fluctuations of the network disorder (randomness of the network structure). 
Ignoring all of the above, and
blindly using Eq.~(\ref{B_UCMF}) with $k_{\min}$$=$$m$, one finds
$\phi_c(\beta)\approx\frac{\gamma+\beta}{\gamma-1} N^{-1}$,
which is a monotonically increasing function of $\beta$ in the semi-infinite region $\beta$$<$$-1$. 
Thus, within our crude approximate scheme, the throughput is maximum at $\beta^*$$=$$-1$.

\begin{figure}[t]
\centering
\vspace*{1.80truecm}
       \includegraphics{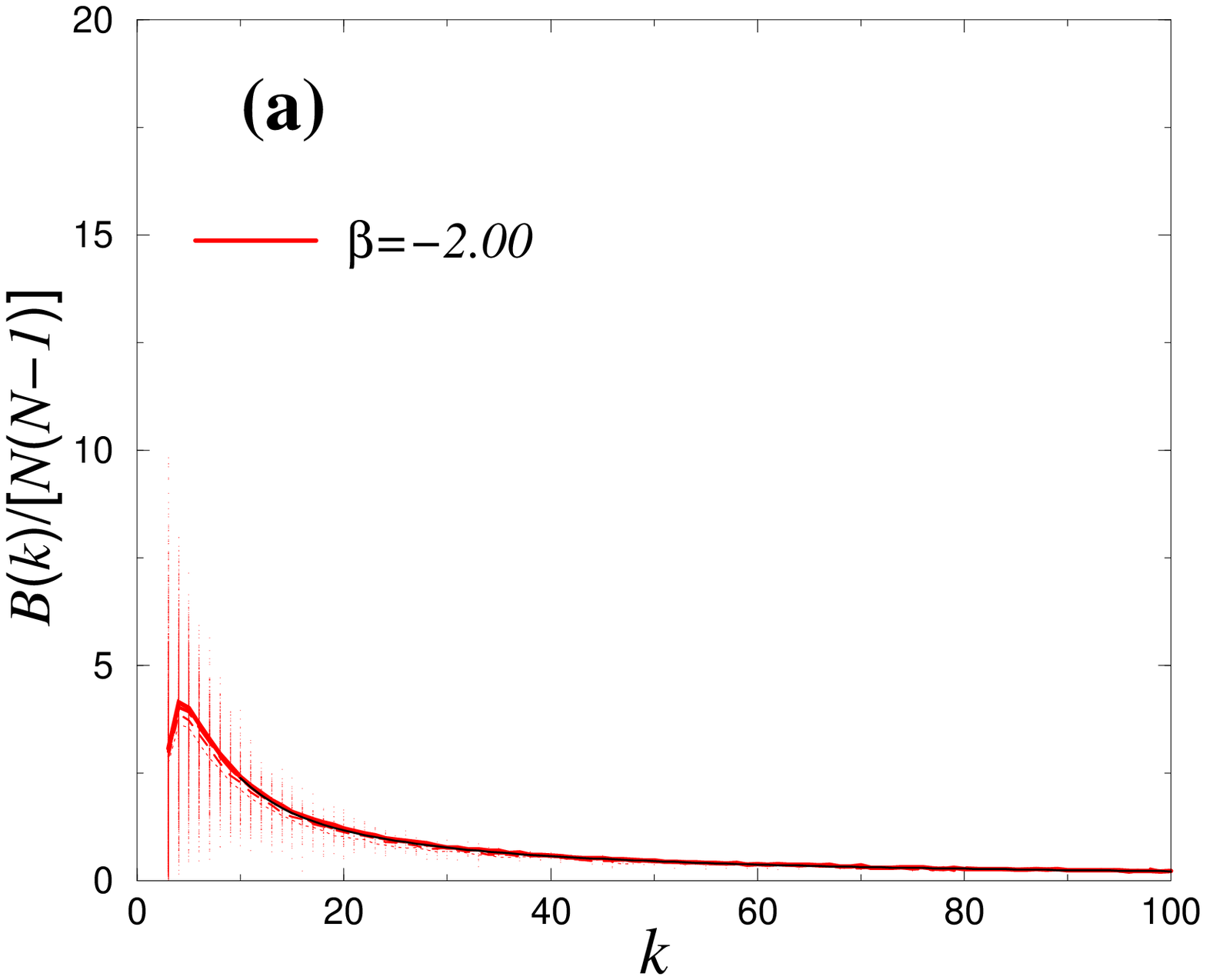}
       \includegraphics{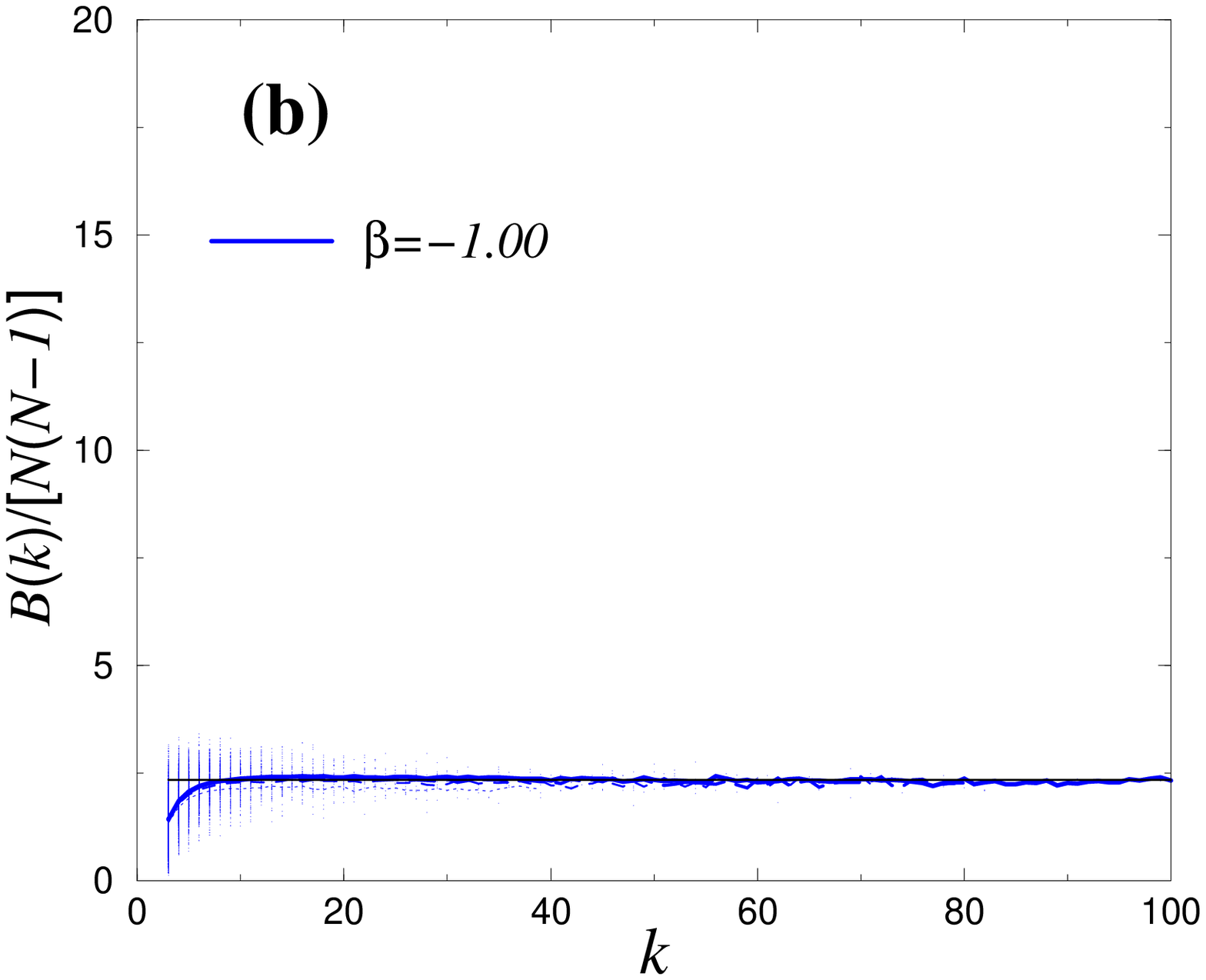}
\vspace*{5.00truecm}
       \includegraphics{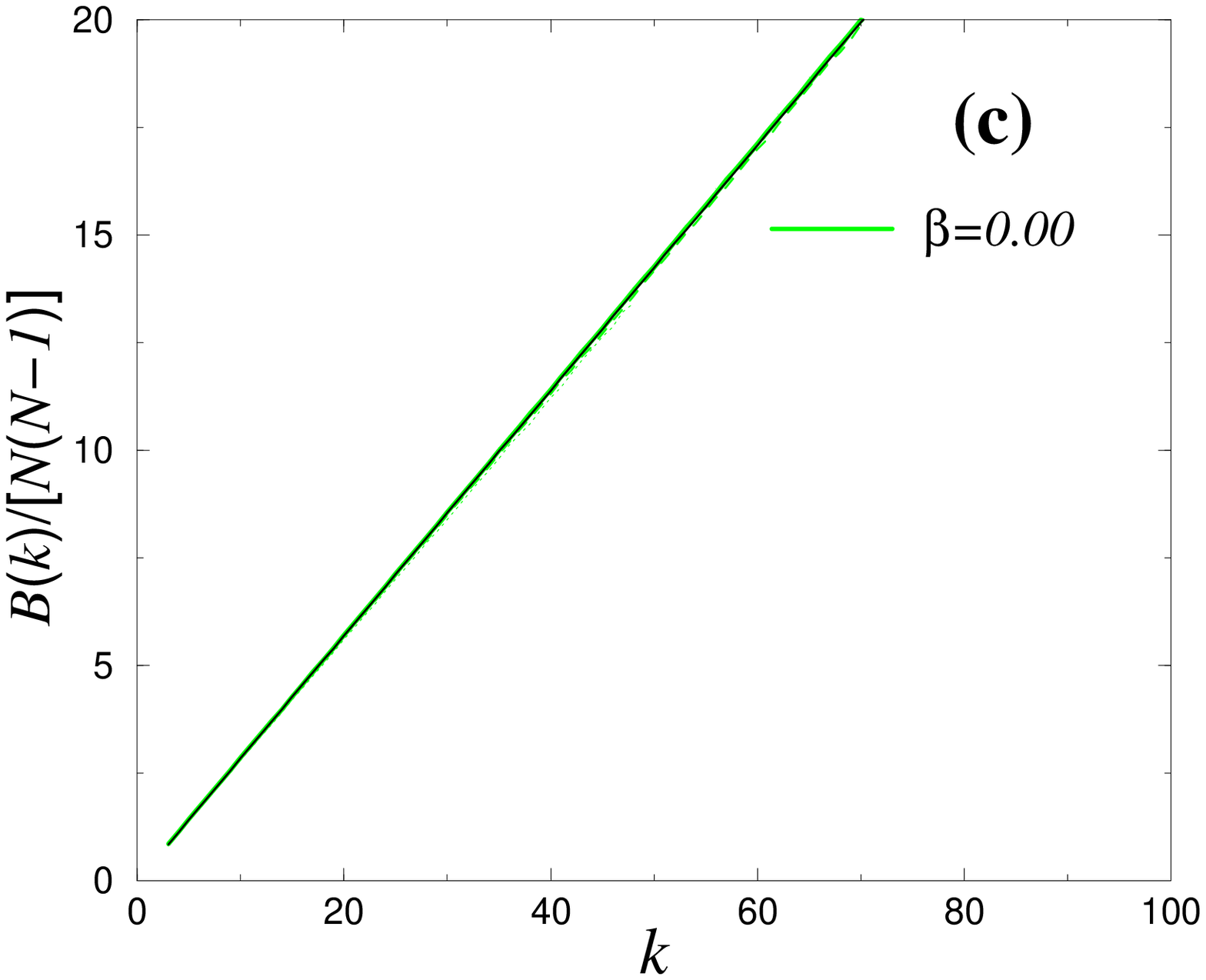}
       \includegraphics{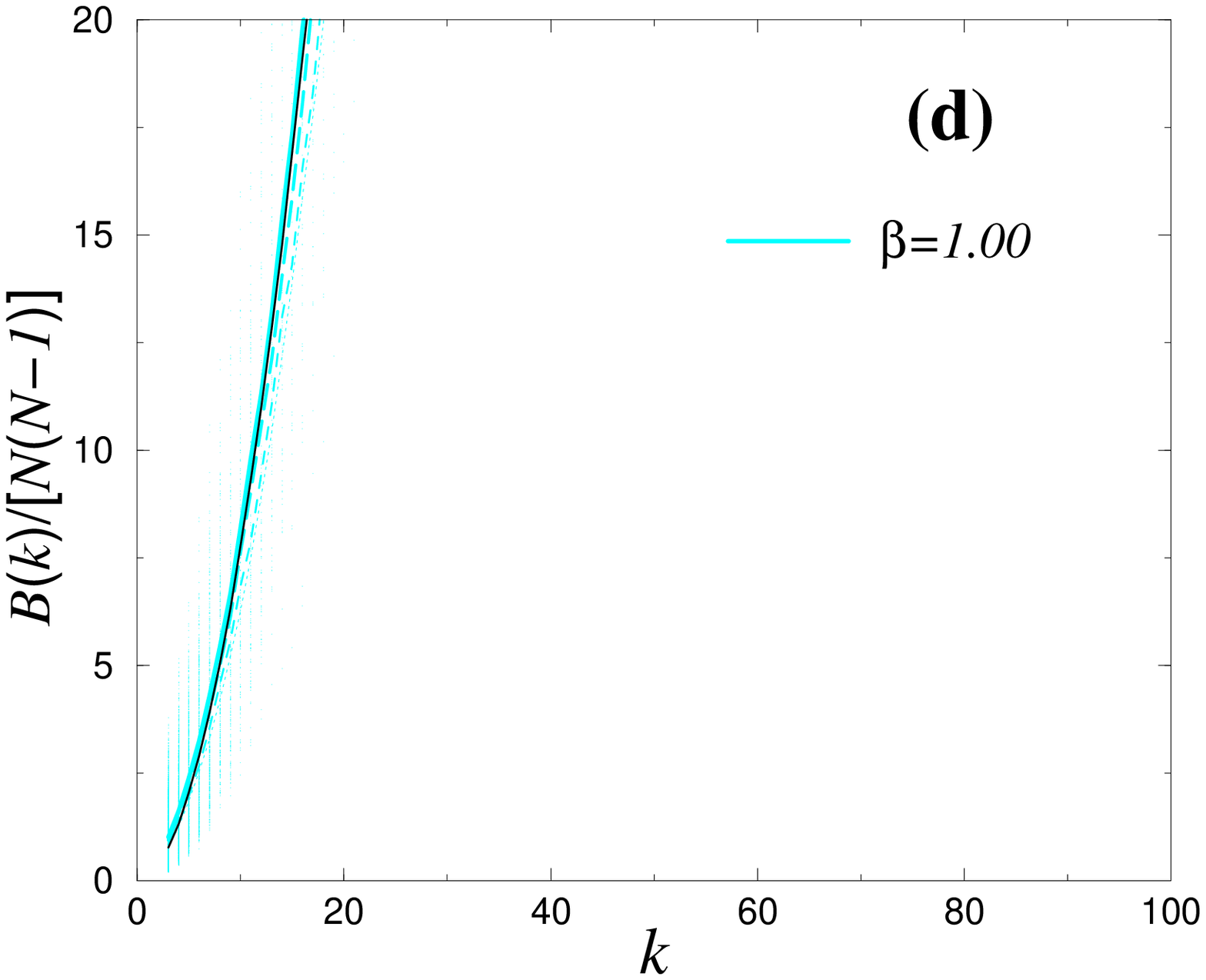}
\vspace*{3.50truecm}
\caption{Normalized RW node betweenness on BA networks with $m$$=$$3$
       as a function of the degree of the nodes for four system sizes
       [$N$$=$ 200 (dotted), 400 (dashed), 1000 (long-dashed), 2000 (solid)]
       for (a) $\beta$$=$$-2.00$, (b) $\beta$$=$$-1.00$, (c) $\beta$$=$$0.00$, and (d) $\beta$$=$$1.00$.
       Data point represented by lines are averaged over all nodes with
       degree $k$. Data for different system sizes are essentially indistinguishable.
       Scatter plot (dots) for the individual nodes is also shown from ten
       network realizations for $N$$=$$1000$. Solid curves, corresponding to the MF+UC
       scaling $B(k)\sim k^{\beta+1}$ [Eq.~(\ref{B_UCMF})], are also shown.}
\label{fig4}
\end{figure}

\begin{figure}[t]
\centering
\vspace*{3.30truecm}
       \includegraphics{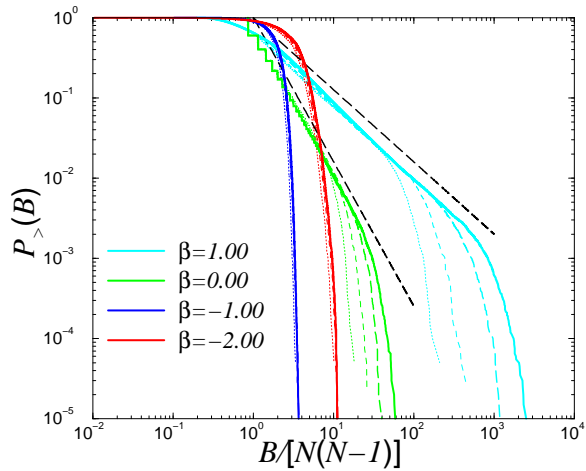}
\vspace*{3.50truecm}
\caption{Cumulative distribution of the normalized RW node betweenness
       for BA networks with $m$$=$$3$ for four system sizes
       [$N$$=$ 200 (dotted), 400 (dashed), 1000 (long-dashed), 2000
       (solid curves)], and for various values of $\beta$ indicated
       in the figure. Straight dashed lines correspond to the
       predicted power-law tail of the cumulative distribution for $\beta$$>$$-1$,
       $P_{>}(B)\sim B^{(1-\gamma)/(1+\beta)}$.}
\label{fig5}
\end{figure}

Numerical work, performed on the BA network $(\gamma$$=$$3$), supports the above
simple analysis. The BA network is somewhat special, in that
correlations are anomalously weak (or marginal), and the structural
and natural cutoffs exhibit the {\em same} ${\cal O}(N^{1/2})$ scaling with
the system size. Testing our MF+UC predictions, we find that
the betweenness is, indeed, strongly correlated with the
degree, in line with Eq.~(\ref{B_UCMF}) [Fig.~\ref{fig4}].
Further, for $\beta>\beta^{*}$$\approx$$-1$, the tail of the degree distribution
governs the tail of the distribution of the
betweenness. Specifically, the cumulative degree distribution,
$P_{>}(k)\sim k^{1-\gamma}$ translates to the cumulative betweenness
distribution $P_{>}(B)\sim B^{(1-\gamma)/(1+\beta)}$ [Fig.~\ref{fig5}].
For $\beta<\beta^{*}$$\approx$$-1$, as noted above, the large-$B$ tail of the betweenness
distribution is coming from the small-$k$ behavior of the degree
distribution. While there is a strict lower cutoff in the degrees $m$, there are many nodes
with degree $m$. It is then the quenched randomness in the particular
network realization which ultimately governs the upper cutoff of the betweenness 
(through the weighted degrees $C_i$).
The tail of the betweenness distribution is essentially independent of
$N$ and numerically found to scale in an exponential-like fashion [Fig.~\ref{fig5}]. 

\begin{figure}[t]
\centering
\vspace*{2.3truecm}
       \includegraphics{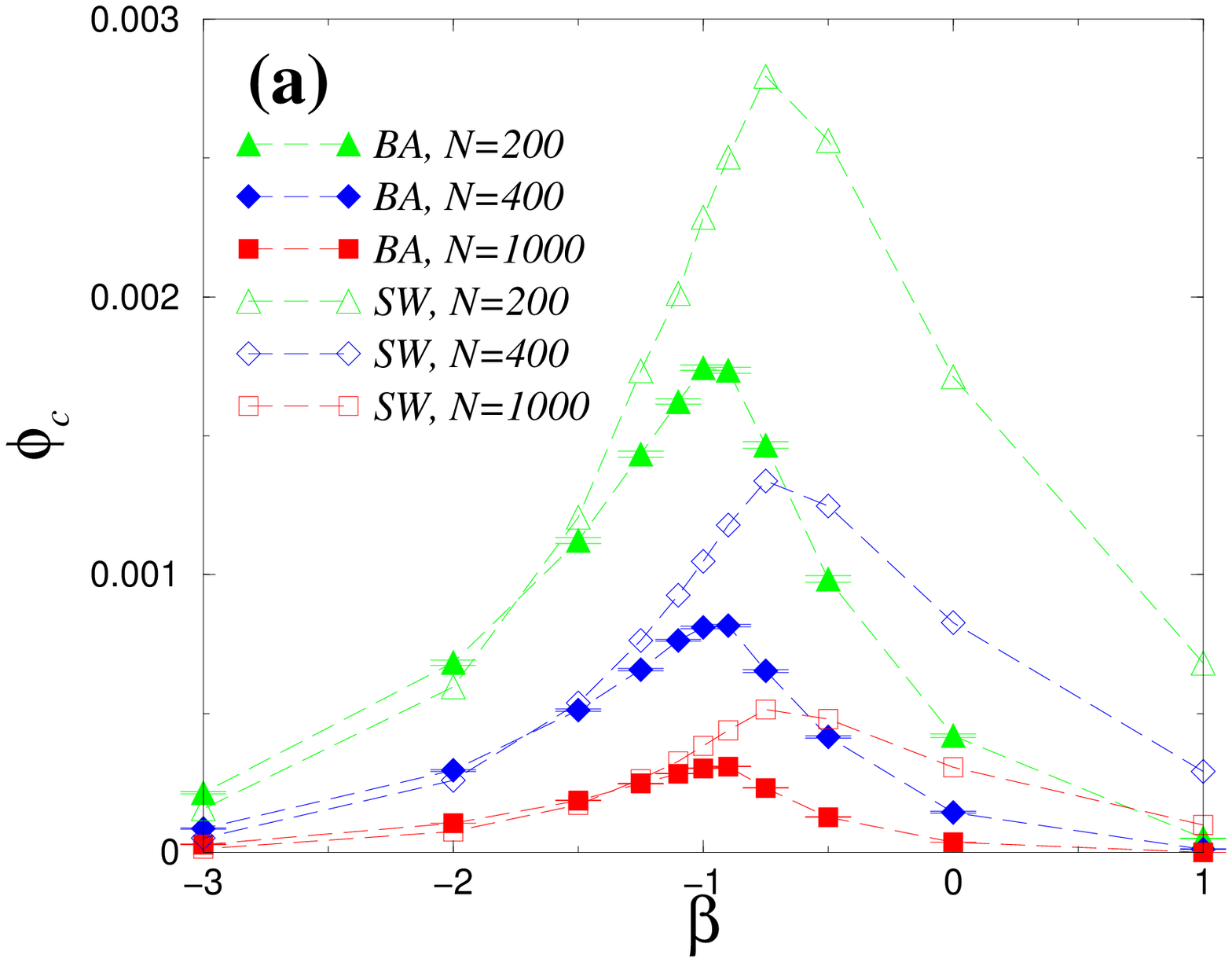}
       \includegraphics{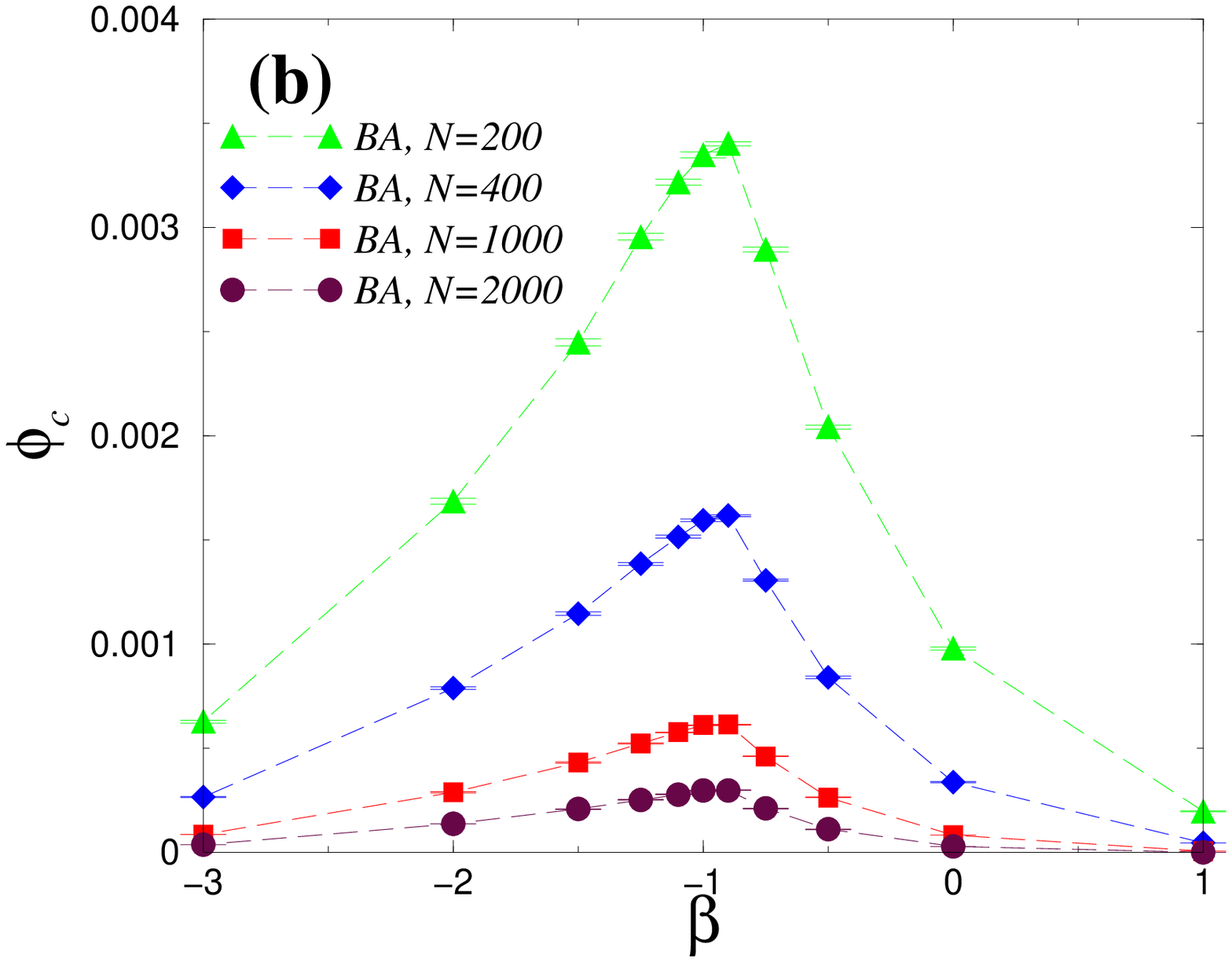}
\vspace*{3.50truecm}
\caption{Critical network throughput per node as a function of the
         weighting parameter $\beta$ for BA networks (solid symbols) for various system
         size for (a) $m$$=$$3$ and for (b) $m$$=$$10$. Figure (a) also shows
         the same observable for SW networks (the same respective open
         symbols) for the same system sizes.}
\label{fig6}
\end{figure}

\begin{figure}[t]
\centering
\vspace*{2.3truecm}
       \includegraphics{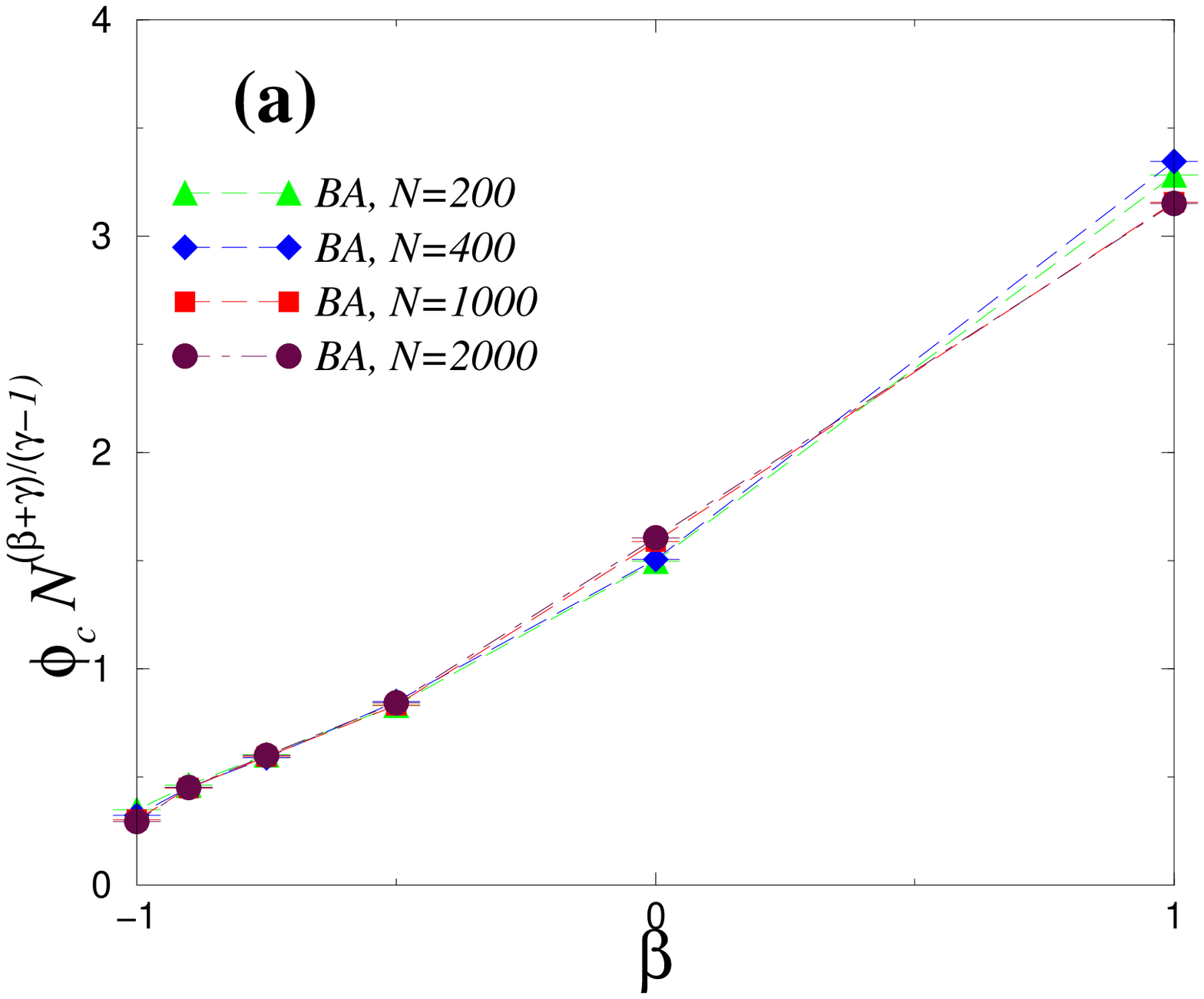}
       \includegraphics{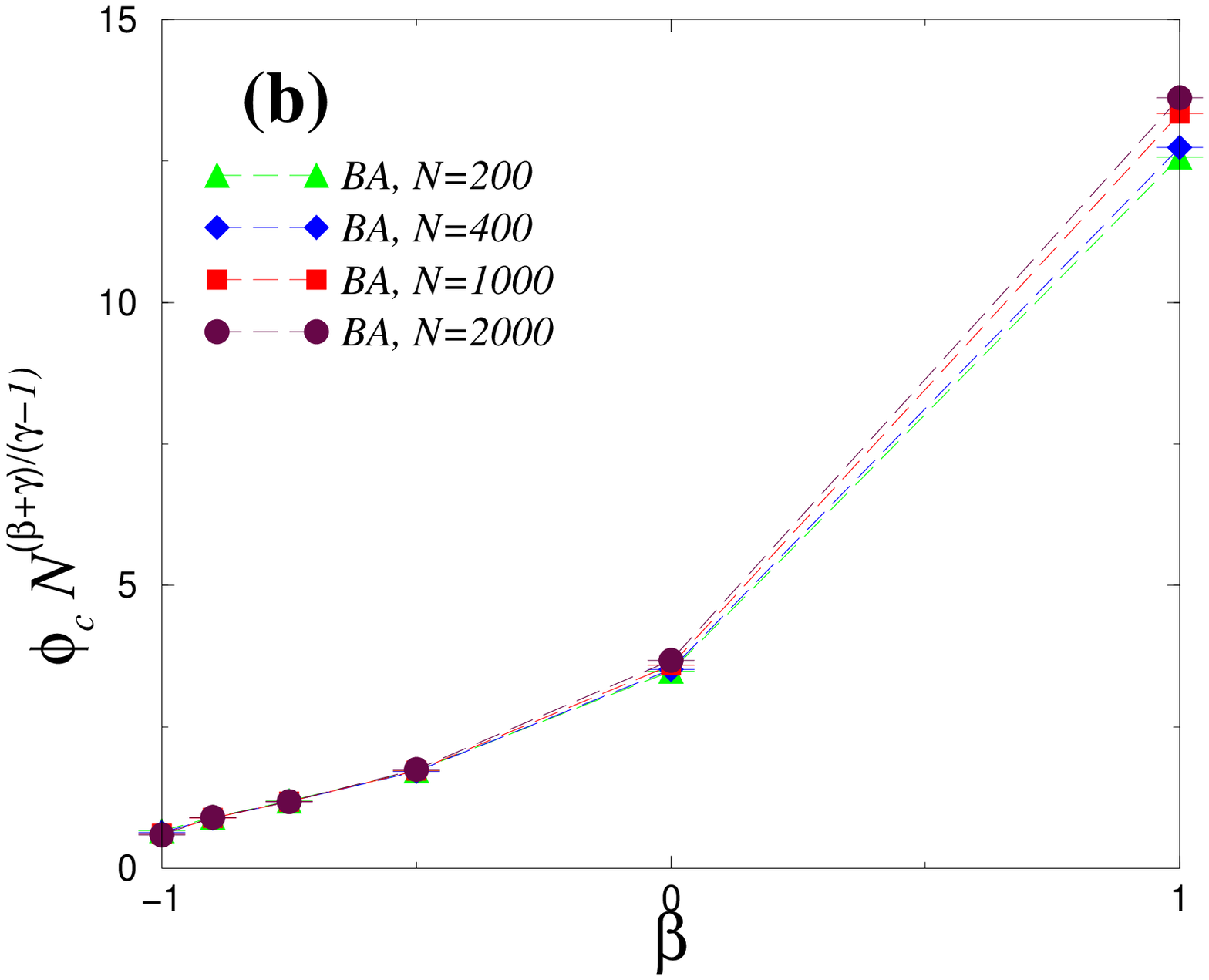}
\vspace*{3.50truecm}
\caption{Scaled critical network throughput per node on BA networks as
       a function of the weighting parameter $\beta$ in the
       $\beta$$>$$-1$ region for (a) $m$$=$$3$ and  (b) $m$$=$$10$.}
\label{fig7}
\end{figure}

As qualitatively predicted by the MF+UC approximation, the critical network
throughput $\phi_c(\beta)$ exhibits a maximum at around
$\beta^{*}$$\approx$$-1$, corresponding to the optimal
weighting scheme, as shown in Figs.~\ref{fig6}.
Further, in the $\beta$$>$$-1$ region, where the long tail of the degree distribution
dominates the network behavior, the network throughput
scales with the number of nodes as $\sim N^{-(\gamma+\beta)/(\gamma-1)}$. [Note that for
the BA network ($\gamma$$=$$3$), the scaling with $N$ by Eqs.~(\ref{phi_c_CM})
and (\ref{phi_c_UCM}) coincide.] The results for the scaled throughput
are shown in Figs.~\ref{fig7}.

In a recent, more realistic network traffic simulation study of a
congestion-aware routing scheme, Danila et al. \cite{Danila_PRE06}
found qualitatively very similar behavior to what we have observed here. In
their network traffic simulation model, packets are forwarded to a neighbor with
a probability  proportional to a power $\beta$ of the {\em instantaneous queue length} of
the neighbor. They found that there is an optimal value of the
exponent $\beta$, close to $-1$.

We also show numerical results for network throughput for SW networks
with the same degree [Fig.~\ref{fig6}(a)]. In particular, an optimally
weighted SW network always outperforms its BA scale-free counterpart
with the same degree. Qualitatively similar results have been obtained in actual traffic simulation 
for networks with exponential degree distribution \cite{Danila_PRE06}.

To summarize, the above simple weighted RW model for local routing on
SF networks indicates that the routing scheme is optimal around the value
$\beta^{*}$$\approx$$-1$. At this point, the load is balanced
[Eq.~(\ref{B_UCMF}) and Fig.~\ref{fig4}(b)], both the average load and
the average packet delivery time are minimum, and the network throughput is
maximum [Fig.~\ref{fig6}].

From a viewpoint of network vulnerability
\cite{BarabERROR,Havlin00,Havlin01,Moreno_EPL2003,ABBV_2006}, the above results for the
weighted RW routing scheme also implies the following.
Network failures are often triggered by large load fluctuations at
a specific node, then subsequently cascading through the system \cite{Moreno_EPL2003}.
Consider a ``normal'' operating scenario (i.e., failure
is {\em not} due to intentional/targeted attacks), where one gradually increases the
packet creation rate $\phi$ and the overloaded nodes (ones with the
highest betweenness) gradually removed from the network \cite{ABBV_2006}. For
$\beta>\beta^{*}$$\approx$$-1$ (including the unweighted RW with
$\beta$$=$$0$), these nodes are the ones with the
highest degrees, but uncorrelated SF networks are structurally vulnerable to removing the hubs.
At the optimal value of $\beta$, not only the network throughput is
maximal, and the average packet delivery time is minimal, but the load
is balanced: overloads are essentially equally likely to occur at any
node and the underlying SF structure is rather resilient to random node
removal \cite{BarabERROR, Havlin00}. Thus, at the optimal value of $\beta$, the local weighted RW
routing simultaneously optimizes network performance and makes the network less
vulnerable against inherent system failures due to congestions at the
processing nodes.

\section{Summary}

We studied the EW process, a prototypical
synchronization problem in noisy environments, on weighted
uncorrelated scale-free networks. 
We considered a specific form of the weights, where the
strength (and the associated cost) of a link is proportional to
$(k_{i}k_{j})^{\beta}$ with $k_{i}$ and $k_{j}$ being the
degrees of the nodes connected by the link. Subject to the constraint
that the total network cost is fixed, we found that in the mean-field
approximation on uncorrelated scale-free graphs, synchronization is optimal
at $\beta^{*}$$=$$-1$. Numerical results, based on exact numerical
diagonalization of the corresponding network Laplacian on BA SF networks, confirmed the
mean-field results, with small corrections to the optimal value of $\beta^{*}$.
Although here, because of the presence of noise and the cost
constraint, the setup of the problem is quite different, our results are very similar
to that of the synchronization of coupled nonlinear oscillators by Zhou et al. \cite{ZHOU06a}.

Employing our recent connections \cite{Korniss_PLA} between the EW process
and resistor networks, and some well-known connections between random
walks and resistor networks \cite{Doyle,Lovasz,Redner,Chandra,Tetali},
we also explored a naturally related problem of weighted random
walks. 
For the simple toy problem, we found that using the associated RW transition
probabilities proportional to a power $\beta$ of the degree of the neighbors [Eq.~(\ref{P_ij_spec})], the
local ``routing'' is optimal when the $\beta^{*}$$=$$-1$ (in the
mean-field approximation). At this optimal network operation point, the load is
balanced, both the average load and the average packet delivery time
are minimum, and the network throughput is maximum. Since the load is
balanced, and thus, can lead to local overloads and subsequent failures at any nodes with
roughly equal probabilities, the above optimal operating point is also the
most resilient one for the underlying scale-free communication network.
While the above local weighted ``routing'' is overly simplified, some aspects of it
can be possibly combined with existing 
realistic protocols to optimize performance in queue-limited communication
networks. For example, existing protocols often utilize an appropriately defined metric for each
node, capturing their ``distance''  (the number of hops) to the
current target \cite{Ye,Chen}. A node then forwards the packet to a neighbor, which is
closer to the target than itself. There may be many nodes satisfying
this criterion, so the forwarding node could employ the weighting RW scheme 
[Eq.~(\ref{P_ij_spec})], applied to this subset, to select the next
node. This may result in improved delivery times and in the 
delaying of the onset of congestion.

\ack
Discussions with K.E. Bassler, B. Kozma, Q. Lu, B.K. Szymanski,
M.B. Hastings, M.J. Berryman, and D. Abbott are gratefully
acknowledged. The author wishes to thank T. Caraco, Z. R\'acz, and M.A. Novotny for a
careful reading of the manuscript. This research was supported in part
by NSF Grant No. DMR-0426488 and RPI's Seed Grant.


\end{document}